\documentclass[twocolumn]{aastex62}

\graphicspath{{./}{figures/}}
\usepackage[utf8]{inputenc}
\usepackage{graphicx}% Include figure file
\usepackage{dcolumn}% Align table columns on decimal point
\usepackage{bm}% bold math
\usepackage{ulem} %here for \sout, removed to avoid funny underlines
\usepackage{hyperref}% add hypertext capabilities
\usepackage{multirow,enumitem}
\usepackage{amsmath}
\usepackage{color}
\usepackage{xcolor}
\usepackage{multirow}

\definecolor{amethyst}{rgb}{0.6, 0.4, 0.8}
\definecolor{pink}{rgb}{0.8, 0.2, 0.8}

\begin{document}

\title{Revisiting Metastable Dark Energy and Tensions in the Estimation of Cosmological Parameters}
\correspondingauthor{Arman Shafieloo}
\email{shafieloo@kasi.re.kr}

\author{Xiaolei Li}
\affiliation{Department of Physics, Hebei Normal University, Shijiazhuang 050024, China}
\affiliation{Korea Astronomy and Space Science Institute, Daejeon 34055, Korea}
\affiliation{Quantum Universe Center, Korean Institute of Advanced Studies, Hoegiro 87, Dongdaemun-gu, Seoul 130-722, Korea}
\author{Arman Shafieloo}
\affiliation{Korea Astronomy and Space Science Institute, Daejeon 34055, Korea}
\affiliation{University of Science and Technology, Yuseong-gu 217 Gajeong-ro, Daejeon 34113, Korea}
\author{Varun Sahni}
\affiliation{Inter-University Centre for Astronomy and Astrophysics, Pune, India}
\author{Alexei A. Starobinsky}

\affiliation{L. D. Landau Institute for Theoretical Physics, Moscow 119334, Russia}
\affiliation{National Research University Higher School of Economics, Moscow 101000, Russia}

% \author{Xiaolei Li}

\date{\today}% It is always \today, today,
 
\begin{abstract}

%We investigate observational constrains on metastable dark energy (DE) models with so far the largest Type \uppercase\expandafter{\romannumeral1}a Supernovae (SNe \uppercase\expandafter{\romannumeral1}a) data set -- Pantheon sample -- in combination with the latest baryonic acoustic oscillations (BAO) data from 6dF Galaxy Survey (6dFGS), the SDSS DR7 main galaxies sample (MGS), the BOSS DR12 galaxies, the eBOSS DR14 quasars as well as  high redshift measurement from complete SDSS-III Ly$\alpha$-quasar cross-correlation function at $z=2.4$ and Cosmological Microwave Background radiation (CMB) distance priors from the finally released \textit{Planck} data in 2018. 
We investigate constraints on some key cosmological parameters by confronting metastable dark energy models with different combinations of the most recent cosmological observations. Along with the standard $\Lambda$CDM model, two phenomenological metastable dark energy models are considered: (\romannumeral1) DE decays exponentially, (\romannumeral2) DE decays into dark matter. We find that: (1) when considering the most recent supernovae and BAO data, and assuming a fiducial $\Lambda$CDM model, the inconsistency in the estimated value of the $\Omega_{\rm{m,0}}h^2$ parameter obtained by either including or excluding Planck CMB data becomes very much substantial and points to a clear tension~\citep{sahni2014model,zhao2017dynamical}; (2) although the two metastable dark energy models that we study provide greater flexibility in fitting the data, and they indeed fit the SNe Ia+BAO data substantially better than $\Lambda$CDM, they are not able to alleviate this tension significantly when CMB data are included; (3) while local measurements of the Hubble constant are significantly higher relative to the estimated value of $H_0$ in our models (obtained by fitting to SNe Ia and BAO data), the situation seems to be rather complicated with hints of inconsistency among different observational data sets (CMB, SNe Ia+BAO and local $H_0$ measurements). Our results indicate that we might not be able to remove the current tensions among different cosmological observations by considering simple modifications of the standard model or by introducing minimal dark energy models. A complicated form of expansion history, different systematics in different data and/or a non-conventional model of the early Universe might be responsible for these tensions. 

%The metastable DE models analyzing with current data set is not able to %reduce the $H_0$ tension between CMB and high redshift BAO measurement from Ly$\alpha$. (2) Adding BAO measurement from Ly$\alpha$ makes $\Gamma\,>\,0$, which means that the DE density is decreasing. Including CMB measurements into the analysis gives $\Gamma\,\simeq\,0$, which shows consistence with $\rm{\Lambda}$CDM model. 
% (3) The degeneracy of contours for $\Omega_{\rm{m,0}}$ vs $H_0$ for model II is different from  previous analysis.

\end{abstract}


\section{Introduction}
The nature of Dark energy (DE) is a key open issue in modern cosmology.
The presence of DE may be required to explain an accelerating universe  as suggested by observations of Type \uppercase\expandafter{\romannumeral1}a Supernovae (SNe \uppercase\expandafter{\romannumeral1}a) \citep{riess1998observational,perlmutter1999measurements}, and supported by measurements of large scale structure (LSS) and the cosmic microwave background radiation (CMB) \citep{spergel2003first,abazajian2004second,tegmark2004cosmological,eisenstein2005detection,komatsu2009five}. The simplest and best known candidate for dark energy is the cosmological constant $\Lambda$  
whose value remains unchanged as the Universe expands. While current observational data are in agreement with the standard $\rm{\Lambda CDM}$ cosmology, General Relativity (GR) with a cosmological constant, though being completely consistent intrinsically at the classical level and having no more problems than GR itself at the quantum level, faces some well-known theoretical difficulties, such as the “fine-tuning” and “cosmic coincidence” problems, when trying to relate the observed small and positive value of the cosmological constant to parameters of the Standard Model of elementary particles and its generalizations like the string theory \citep{sahni2000case,bean2005insights}. Recent papers have also drawn attention to some other difficulties faced by $\rm{\Lambda CDM}$ including tension between the value of $H_0$ estimated by fitting to the acoustic peaks in the \textit{Planck} CMB power spectrum \citep{collaboration2014planck,ade2016planck,aghanim2018planck} and that obtained from distance scale estimates \citep{sahni2014model,ding2015there,zheng2016omh,sola2017h0,alam2017constraining,shanks2018gaia}.  {{In order to alleviate these problems, different solutions such as  dark energy models beyond $\Lambda$CDM model, modifications to general relativity theory and  other physically-motivated possibilities like modifications to the dark matter sector have been put forward \citep{ko2017hidden,raveri2017partially,kumar2017echo,di2017can,renk2017galileon,sola2017h0,di2018vacuum,khosravi2019h,poulin2019early,vattis2019dark,li2019phenomenologically,pan2019reconciling}. }}

%Moreover, an interaction between cosmic dark sectors has been considered \citep{olivares2006matter,wang2016dark,zheng2017ultra} and they find that the interaction DE models can avoid the coincidence problem and alleviate $H_0$ tension. 

A new class of `Metastable DE' models was introduced in \citet{shafieloo2017metastable}. In these models DE decays into other dark sector components of the Universe such as dark matter or dark radiation. The rate of decay of DE depends only upon its `intrinsic' properties and not on extrinsic considerations such as the rate of expansion of the universe, etc. The metastable DE model was largely inspired by the radioactive decay of heavy nuclei into lighter elements.
A total of three  metastable DE models were considered, namely, \romannumeral1) DE decays exponentially, \romannumeral2) DE decays into non-baryonic Dark Matter (DM), \romannumeral3) DE decays into Dark Radiation. {{We should note that from a theoretical perspective one can achieve metastable behaviour of dark energy from an intermediate phase of quantum vacuum decay\citep{szydlowski2017quantum,szydlowski2018decay}. }}
It was found that model \uppercase\expandafter{\romannumeral2} showed less tension between CMB and QSO based $H(2.34)$ BAO data than that faced by $\rm{\Lambda CDM}$. 
Clearly in order to understand DE, one has to turn to cosmological observations. In previous work, DE models have been discussed in the context of different kinds of cosmological observations \citep{cao2015cosmology,li2016comparison,zheng2017ultra,shafieloo2017metastable,li2017testing}. For metastable DE models, \citet{shafieloo2017metastable} used  580 SNe Ia from the Union-2.1 compilation \citep{suzuki2012hubble} and four BAO data sets in combination with CMB shift parameters $R$, $l_a$, to place constraints on the DE parameters. Since then more precise data sets have been released.  In this work, we present constraints on two metastable DE models using the SNe Ia Pantheon sample \citep{scolnic2017complete}, latest BAO data from the 6dF Galaxy Survey (6dFGS) \citep{beutler20116df}, the SDSS DR7 main galaxies sample (MGS) \citep{ross2015clustering}, the BOSS DR12 galaxies \citep{alam2017clustering}, newly released eBOSS DR14 \citep{zhao2018clustering} and high redshift measurement from complete SDSS-III Ly$\alpha$-quasar cross-correlation function at $z=2.4$ \citep{des2017baryon} in combination with CMB distance priors from final full-mission \textit{Planck} measurements of the CMB anisotropies \citet{aghanim2018planck,Chen2018pl}. The aim of our analysis is to place constraints on metastable DE models using the latest data, compare metastable DE with $\Lambda$CDM, and check whether the $H_0$ tension has been alleviated.

This paper is organized as follows, in section~\ref{sec:cos_model} we briefly introduce the Friedmann equations for our model. The observational data to be used including SNe Ia, BAO and distance prior from CMB are presented in section \ref{sec:Cosmo_probe}. Section~\ref{sec:res} contains our main results and some discussion. We summarize our results in section~\ref{sec:sum}.

\section{Cosmological models} \label{sec:cos_model}
In this work, we test two metastable DE models: (i) in the first DE decays exponentially, (ii) in the second DE decays into non-baryonic dark matter. For comparison, we also place constraints on $\rm{\Lambda CDM}$ using the same data sets. We assume that the Friedmann - Lema${\rm \hat i}$tre - Robertson - Walker (FLRW) metric is spatially flat that is strongly supported by recent observations \citep{l2017model,shafieloo2018falsifying,aghanim2018planck}. Under this assumption, the angular diameter distance $D_{\rm{A}}(z)$ at redshift $z$ can be written as
\begin{equation}\label{eq:add}
    D_{\rm{A}}(z)\,=\,\frac{c}{H_0(1+z)}\int_0^z \frac{d z'}{E(z')}
\end{equation}
where $E(z)\,=\,H(z)/H_0$ is the expansion rate and $H_0$ is the current value of Hubble parameter. 

\subsection{$\rm{\Lambda CDM}$}
$\rm{\Lambda CDM}$ model is perhaps the simplest of all dark energy models. 
In it the cosmological constant $\Lambda$ plays the role of DE. The Hubble parameter in $\rm{\Lambda CDM}$ has the form 
\begin{equation}
H^2(z)\,=\,H^2_0\left[\Omega_{m,0}(1+z)^3+\Omega_{\rm{DE}} \right]
\end{equation}
where $\Omega_{m,0}$ is the current matter density parameter and $\Omega_{\rm{DE}} = \frac{\Lambda}{3H_0^2}$ is the density parameter associated with dark energy.

\subsection{Model I: Exponentially decaying DE}
In this model, DE decays exponentially as 
\begin{equation}
\dot{\rho}_{\rm{DE}}\,=\,-\Gamma\rho_{\rm{DE}}
\end{equation}
where $\Gamma$ is the only free parameter in this equation and $\Gamma\,>\,0$ or $\Gamma\,<\,0$  means that DE density is decreasing or increasing, respectively. The Hubble parameter obtained from the FRW equations can be written as 
\begin{equation}\label{eq:hz_1}
\begin{aligned}
H^2(z)\, = & \,H_0^2 \left[ {\Omega_{m,0}(1+z)^3} \right. \\
& +\left.(1-\Omega_{m,0}) {\rm{exp}}\left(\frac{\Gamma}{H_0}\int_0^z\frac{dz'}{E(z')(1+z')}\right)\right] 
\end{aligned}
\end{equation}

\subsection{Model II: DE decays into DM}
In this model dark energy decays into non-baryonic dark matter as follows:
\begin{equation}
\dot{\rho}_{\rm{DE}}\,=\,-\Gamma \rho_{\rm{DE}} 
\end{equation}

\begin{equation}
\dot{\rho}_{\rm{DM}}+3H\rho_{\rm{DM}}\,=\,\Gamma \rho_{\rm{DE}} 
\end{equation}

This model is effectively an interacting DE-DM model since when $\Gamma\,\neq\,0$, energy is exchanged between DM and DE. The Hubble parameter for this model can be written as
\begin{equation}
    H^2(z)\,=\,H_0^2\left[\Omega_{\rm{DE}}(z)+\Omega_{\rm{DM}}(z)+\Omega_{b,0}(1+z)^{3}\right]
\end{equation}
Here $\Omega_{b,0}$ is the baryon density. Since in this model DE interacts with non-baryonic DM, we need to separate DM density from baryon density. While for metastable DE model I, dark matter and baryon matter can be treated as a whole, e.g., $\Omega_{\rm{m,0}}\,=\,\Omega_{\rm{DM,0}}+\Omega_{b,0}$.
%{\bf Varun: If the baryon density is included here then it should be included in the earlier equations also.}

\vspace{0.5cm}
For the metastable DE models, the cosmological parameters to be constrained are $\{\Omega_{m,0}, H_0, \Gamma\}$. Both model I and model II become $\rm{\Lambda CDM}$ when $\Gamma\,=\,0$. We refer the reader to \citet{shafieloo2017metastable} for more details about these two DE models.

\section{Data and analysis}\label{sec:Cosmo_probe}
In this work, we consider the combination of three different kinds of cosmological probes to put constraints on DE models, including SNe Ia as standard candles and BAO together with CMB as standard rulers.

\subsection{Type \uppercase\expandafter{\romannumeral1}a Supernovae}
In their role as standard candles, SNe Ia have been of great importance to measure cosmological distances. In our analysis, we use the new "Pantheon" sample \citep{scolnic2017complete}, which is the largest combined sample of SN Ia and consists of 1048 data with redshifts in the range $0.01\,<\,z\,<\,2.3$.  In order to reduce the impact of calibration systematics on cosmology, the Pantheon compilation used cross-calibration of the photometric systems of all the subsamples used to construct the final sample.

\subsection{Baryonic Acoustic Oscillations} 
The second data set used in our analysis is BAO. It includes lower redshift BAO measurements from galaxy surveys and higher redshift BAO measurement from Lyman-$\alpha$ forest (Ly$\alpha$) data. For the lower redshift BAO observations, we turn to the latest measurements of acoustic-scale distance ratio from the 6-degree Field Galaxy Survey (6dFGS) \citep{beutler20116df}, the SDSS Data Release 7 Main Galaxy sample (MGS) \citep{ross2015clustering}, the BOSS DR12 galaxies \citep{alam2017clustering} and the eBOSS DR14 quasars \citep{zhao2018clustering}, while the higher redshift BAO measurement is derived from the complete SDSS-III Ly$\alpha$ quasar cross-correlation function at $z=2.4$ \citep{des2017baryon}. Details of the BAO measurements are listed in Table~\ref{tab:bao}.

\begin{table*}[!t]
\centering
\caption{BAO measurements used in our analysis. Here $r_{\rm{d}}$ is the comoving sound horizon at the baryon drag epoch $z_{\rm{drag}}$, and $D_{\rm{V}}\,=\,\left[(1+z)^2D^2_{\rm{A}}(z)cz/H(z)\right]^{1/3}$, $D_{\rm{M}}\,=\,(1+z)D_{\rm{A}}$, where $D_{\rm{A}}$ is the angular diameter distance defined in equation~(\ref{eq:add}). The fiducial comoving sound horizon for BOSS DR12, eBOSS DR14 and Ly$\alpha$ is $r_{\rm{d,fid}} = 147.78$ Mpc. In practice, our analysis uses the full covariance matrix for BAO measurements from \citet{alam2017clustering,zhao2018clustering,des2017baryon}.} \label{tab:bao}
\begin{tabular}{cccccc}
\hline
\hline 
 $z$  & $D_{\rm{v}}/r_{\rm{d}}$  &  $D_{\rm{M}}\times(r_{\rm{d,fid}}/r_{\rm{d}})$(Mpc) & $H\times(r_{\rm{d}}/r_{\rm{d,fid}})$(km/s/Mpc) & $D_{\rm{A}}\times(r_{\rm{d,fid}}/r_{\rm{d}})$(Mpc)  &  Ref. \\
 \hline
 0.106 & $3.047\pm0.137$  & - & - & - & \citet{beutler20116df}           \\
 \hline
 0.150 & $4.480\pm0.168$  & - & - & - & \citet{ross2015clustering}           \\
 \hline
 0.38 & - &  $1512\pm24$ & $81.2\pm2.4$ &-&    \\
 0.51 & - &  $1975\pm30$ & $90.9\pm2.4$ &-&  \citet{alam2017clustering}  \\
 0.61 & - &  $2307\pm37$ & $99.0\pm2.5$ & -&    \\
\hline
0.978 & - & - &  $113.72\pm14.63$ & $1586.18\pm284.93$ &  \\
1.230 & - & - &  $131.44\pm12.42$ & $1769.08\pm159.67$ &\citet{zhao2018clustering} \\
1.526 & - & - &  $148.11\pm12.75$ & $1768.77\pm96.59$ & \\
1.944 & - & - &  $172.63\pm14.79$ & $1586.18\pm146.46$ & \\
\hline
 2.4  & - &  $5393.4\pm176.8$ & $227.56\pm5.6$ &-&  \citet{des2017baryon}  \\
\hline
\hline
\end{tabular}
\end{table*}

\subsection{Cosmic Microwave Background}
We include CMB into our analysis by using the CMB distance prior, the acoustic scale $l_{\rm{a}}$ and the shift parameter $R$ together with the baryon density $\Omega_bh^2$. The shift parameter is defined as 
\begin{equation}
R\,\equiv\,\sqrt{\Omega_mH_0^2}r(z_*)/c
\end{equation}
and the acoustic scale is 
\begin{equation}
l_a\,\equiv\,\pi r(z_*)/r_s(z_*)
\end{equation}
where $r(z_*)$ is the comoving distance to the photon-decoupling epoch $z_*$. We use the distance priors from the finally release \textit{Planck} TT, TE, EE +low E data in 2018 \citep{Chen2018pl}, which makes the uncertainties 40\% smaller than those from \textit{Planck} TT+low P.

\vspace{0.5cm}

When using SNe Ia and BAO as cosmological probes, we use a conservative prior for $\Omega_{\rm{b}}h^2$ based on the measurement of D/H by \citet{cooke2018one} and standard BBN with modelling uncertainties. The constraint results are obtained with Markov Chain Monte Carlo (MCMC) estimation using \texttt{CosmoMC} \citep{lewis2002cosmological}.

In our analysis, four kinds of combined data sets are considered: 1) Pantheon compilation in combination with BAO data from 6dFGS, MGS and BOSS DR12. 2) We add BAO data from eBOSS DR14 to the first data set. 3) Adding high redshift BAO measurement from Ly$\alpha$ to the second data combination. 4) Finally, we include the CMB distance prior to the full combination of data sets. 

\section{results and discussion}\label{sec:res}

\begin{figure*}
\centering
\includegraphics[width=0.45\textwidth]{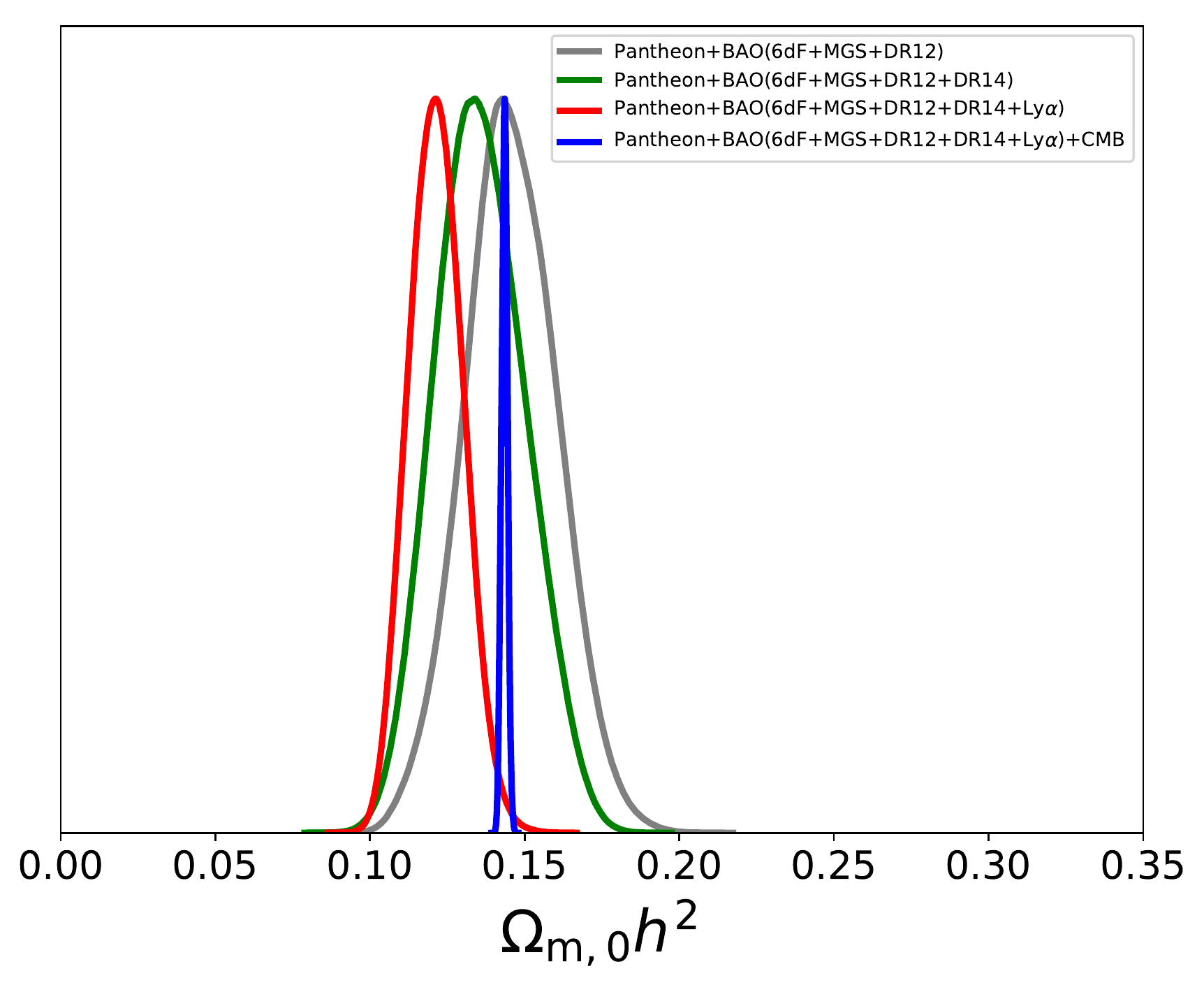}
\includegraphics[width=0.48\textwidth]{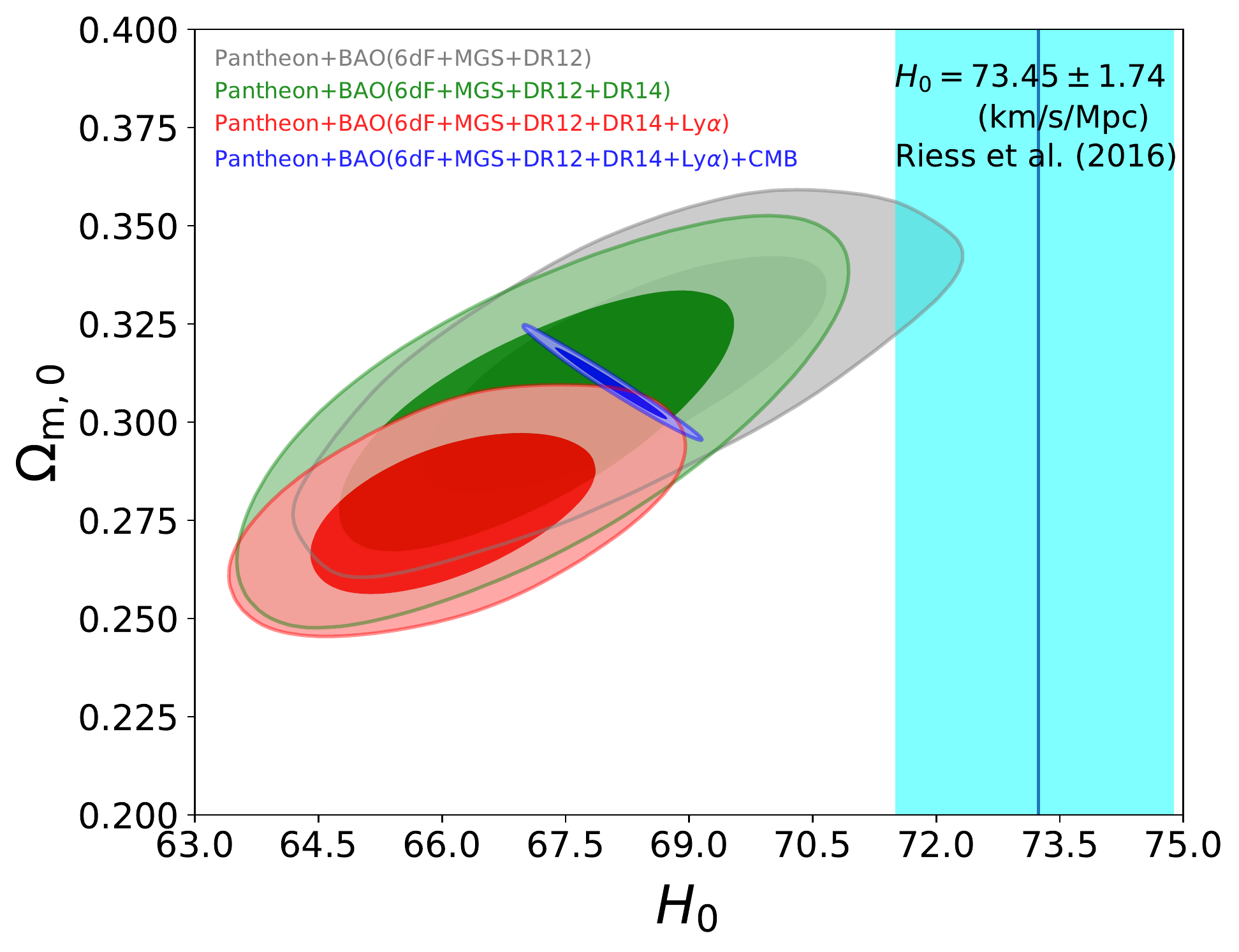}
\caption{Observed constrains on standard $\rm{\Lambda}$CDM. Left plot gives the 1D likelihood for $\Omega_{\rm{m,0}}h^2$ and right plots shows the 1$\sigma$ and 2$\sigma$ contours for $\Omega_{\rm{m,0}}$ vs $H_0$. The cyan shadow in the right plot give $H_0$ results from \citet{riess20162} and we show the constrain results from different data sets in different color.}\label{fig:res_lcdm}
\end{figure*}

\begin{table*}[!t]
\centering
\caption{ The best fit of cosmological parameters (the first row in each parameter row) for {\rm{$\Lambda$}}CDM and its mean value together with its marginalized 1$\sigma$ uncertainties (the second row in each parameter row) as well as their $\chi^2$ value. } \label{tab:res_lcdm}
\begin{tabular}{ccccc}
\hline
\hline 
data       & Pantheon           & Pantheon            & Pantheon             & Pantheon \\
           & +BAO(6dF+MGS+DR12) & +BAO(6dF+MGS+DR12   & +BAO(6dF+MGS+DR12    &+BAO(6dF+MGS+DR12 \\
parameters &                    & +DR14)              &  +DR14+Ly$\alpha$)   & +DR14+Ly$\alpha$)+CMB \\
 \hline
 $\Omega_{m,0}$ & $0.312$                    & $0.299$                   & $0.276$                   &                  $0.310$        \\
                & $0.313^{+0.020}_{-0.021}$  & $0.301^{+0.021}_{-0.020}$ & $0.277^{+0.013}_{-0.014}$ &                  $0.310^{+0.006}_{-0.006}$        \\
 \hline
 $H_0$          & $68.10$                    & $67.03$                   & $66.0$                   &                  $68.06$        \\
                & $68.22^{+1.65}_{-1.69}$    & $67.17^{+1.56}_{-1.54}$   & $66.08^{+1.19}_{-1.14}$ &                  $68.05^{+0.46}_{-0.43}$        \\
 \hline
 $\Omega_{m,0}h^2$ & $0.145$                    & $0.134$                   & $0.1083$                  &                  $0.144$        \\
                   & $0.146^{+0.016}_{-0.015}$  & $0.135^{+0.016}_{-0.013}$ & $0.121^{+0.009}_{-0.009}$ &                  $0.144^{+0.001}_{-0.001}$        \\
 \hline
 $\chi^2$          & 1042.58                   & 1046.94                    &  1064.58                 & 1071.14\\
\hline
\hline
\end{tabular}
\end{table*}

% {\bf{For the completeness of presented statistical analysis, in \citet{li2019phenomenologically,pan2019reconciling}, the authors made comparison of cosmological models to the standard $\Lambda$CDM model.
% Therefore, }}
We first show the constraints for the $\Lambda$CDM model in Fig.~\ref{fig:res_lcdm} where the different colors denote the results from different data combinations. The left plot shows the 1D marginalized results for $\Omega_{m,0}h^2$ and the right plot presents the 2D marginalized 1$\sigma$ and 2$\sigma$ contours of $\Omega_{m,0}$ vs $H_0$. In the right plot, we also show the $H_0$ constraints from \citet{riess20162} in the cyan shadow. The best fit for cosmological parameters of {\rm{$\Lambda$}}CDM and their 1$\sigma$ uncertainties are summarized in Table~\ref{tab:res_lcdm}. We also show the $\chi^2$ from each data combination in Table~\ref{tab:res_lcdm}. From Fig.~\ref{fig:res_lcdm} and Table~\ref{tab:res_lcdm}, we can clearly see that, adding BAO measurements from eBOSS DR14 pushes the best fit of $\Omega_{m,0}$ and $H_0$ towards a lower value (the green curves), and including BAO measurement from Ly$\alpha$ makes the best fit of $\Omega_{m,0}$ and $H_0$ even lower (the red curves). However, a higher matter density and a higher Hubble constant are obtained when using the combined data of Pantheon+BAO(6dF+MGS+DR12+DR14+Ly$\alpha$)+CMB (the blue curves), which makes the best fits of $\Omega_{\rm{m},0}h^2$ in excellent agreement with the results from Pantheon+BAO(6dF+MGS+DR12). One might note that there is a clear tension between the results obtained from adding high redshift Ly$\alpha$ BAO measurement and the results obtained from including CMB. We should note that this tension has been reported earlier by \citet{sahni2014model,shafieloo2017metastable}. It is important to emphasis here that one of our main aims is to see whether we can alleviate this tension by analyzing the metastable DE models using current data sets.
\vspace{0.5cm}

\begin{figure*}
\centering
\includegraphics[width=0.45\textwidth]{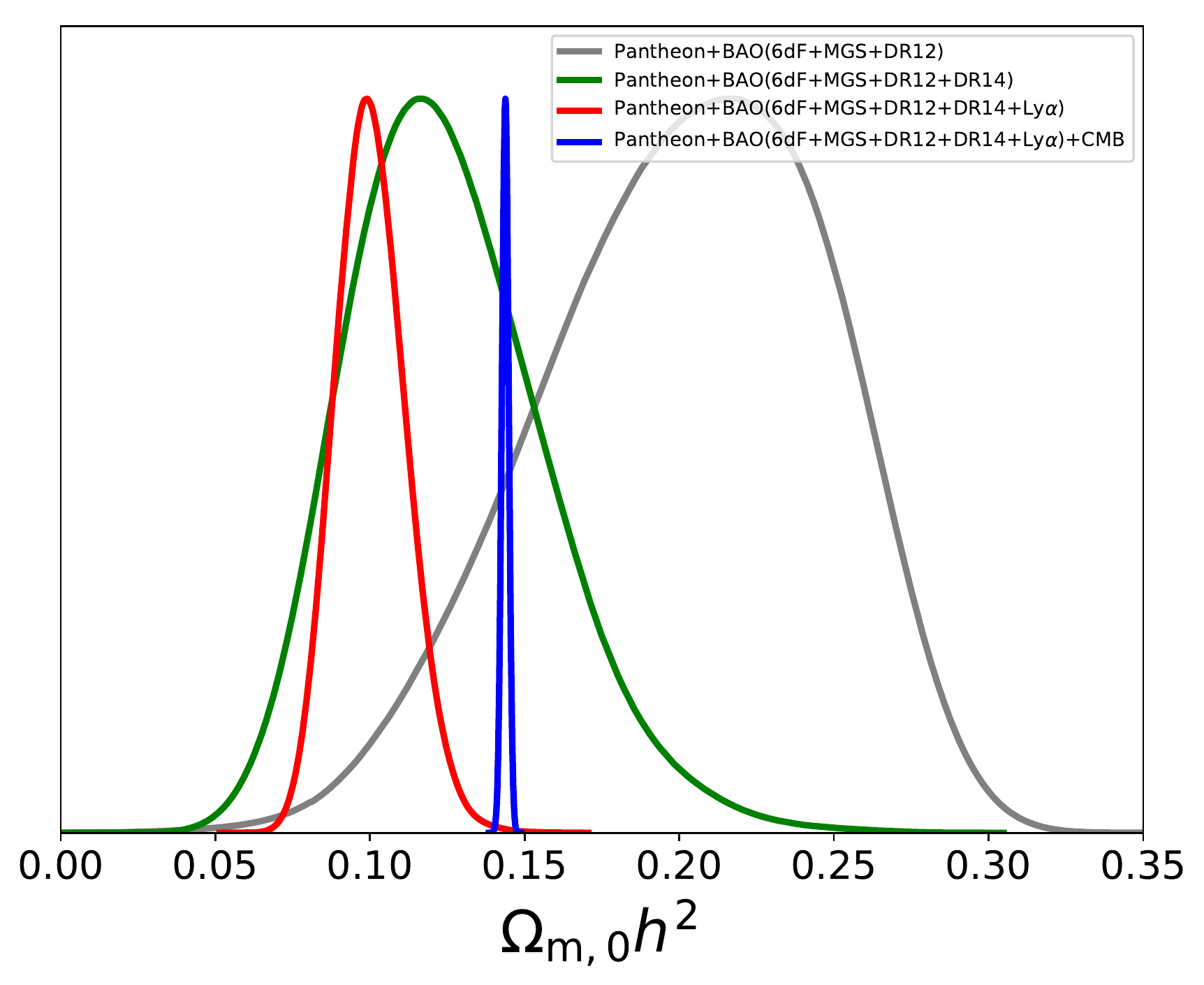}
\includegraphics[width=0.45\textwidth]{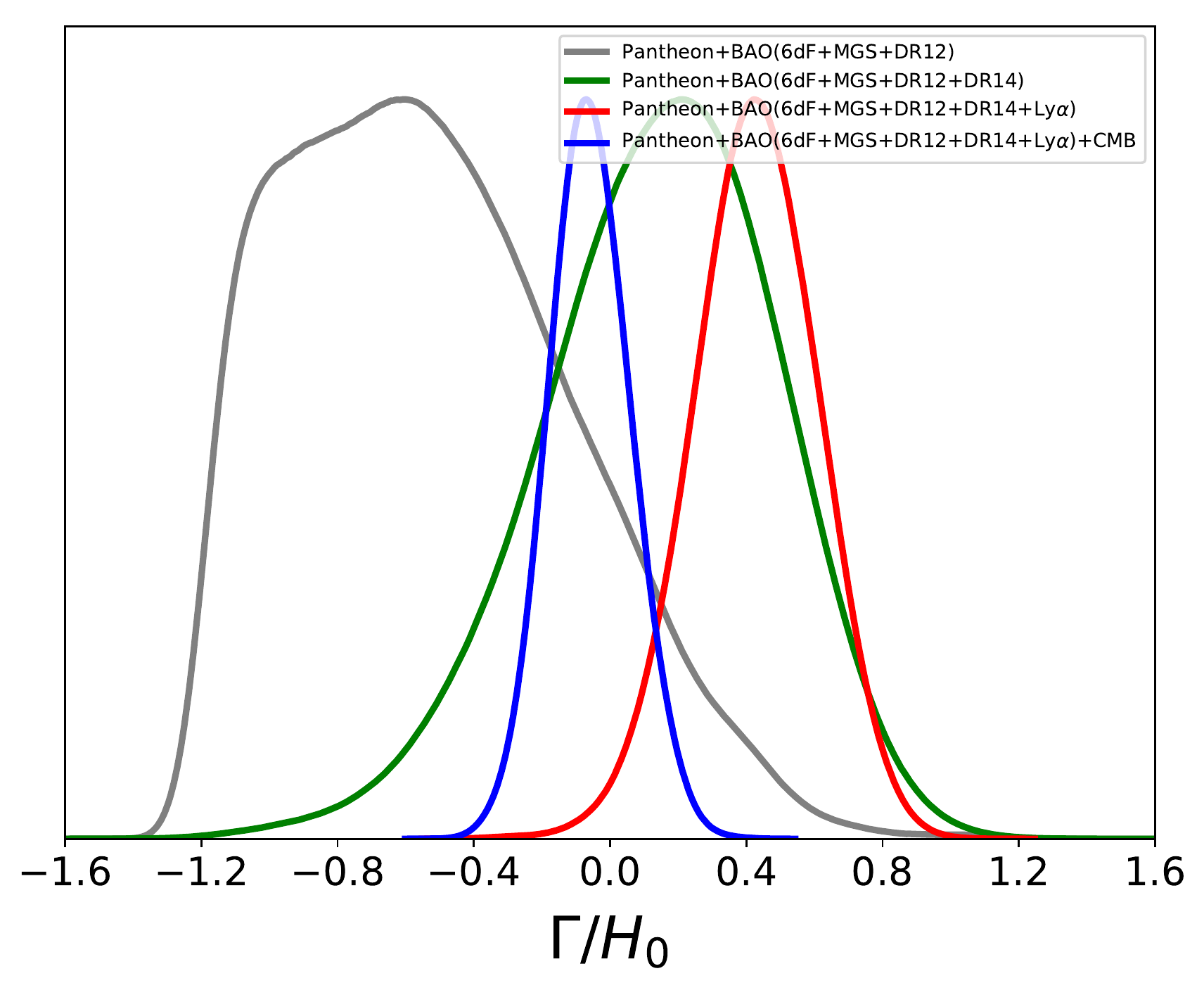}
\includegraphics[width=0.47\textwidth]{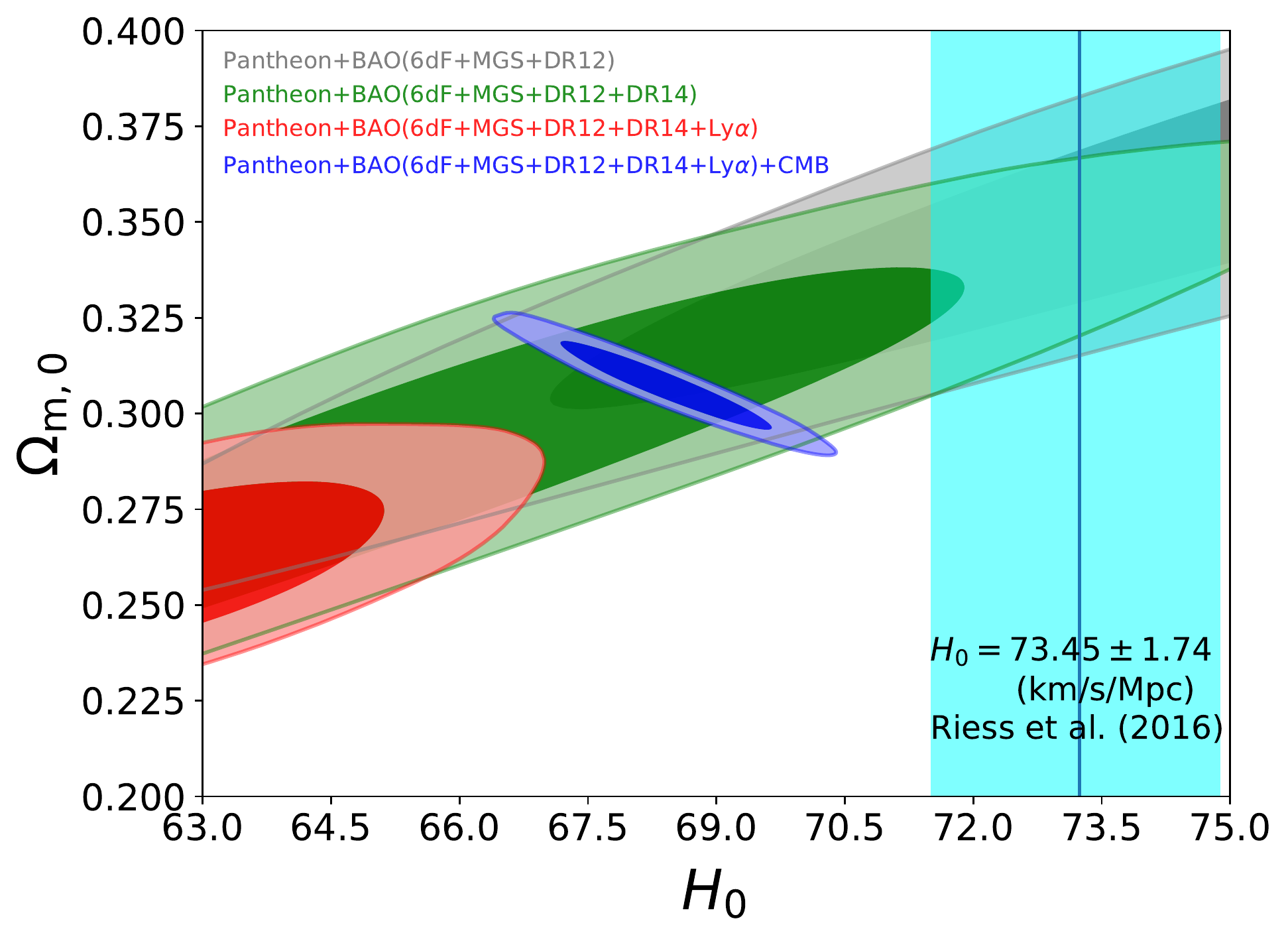}
\includegraphics[width=0.45\textwidth]{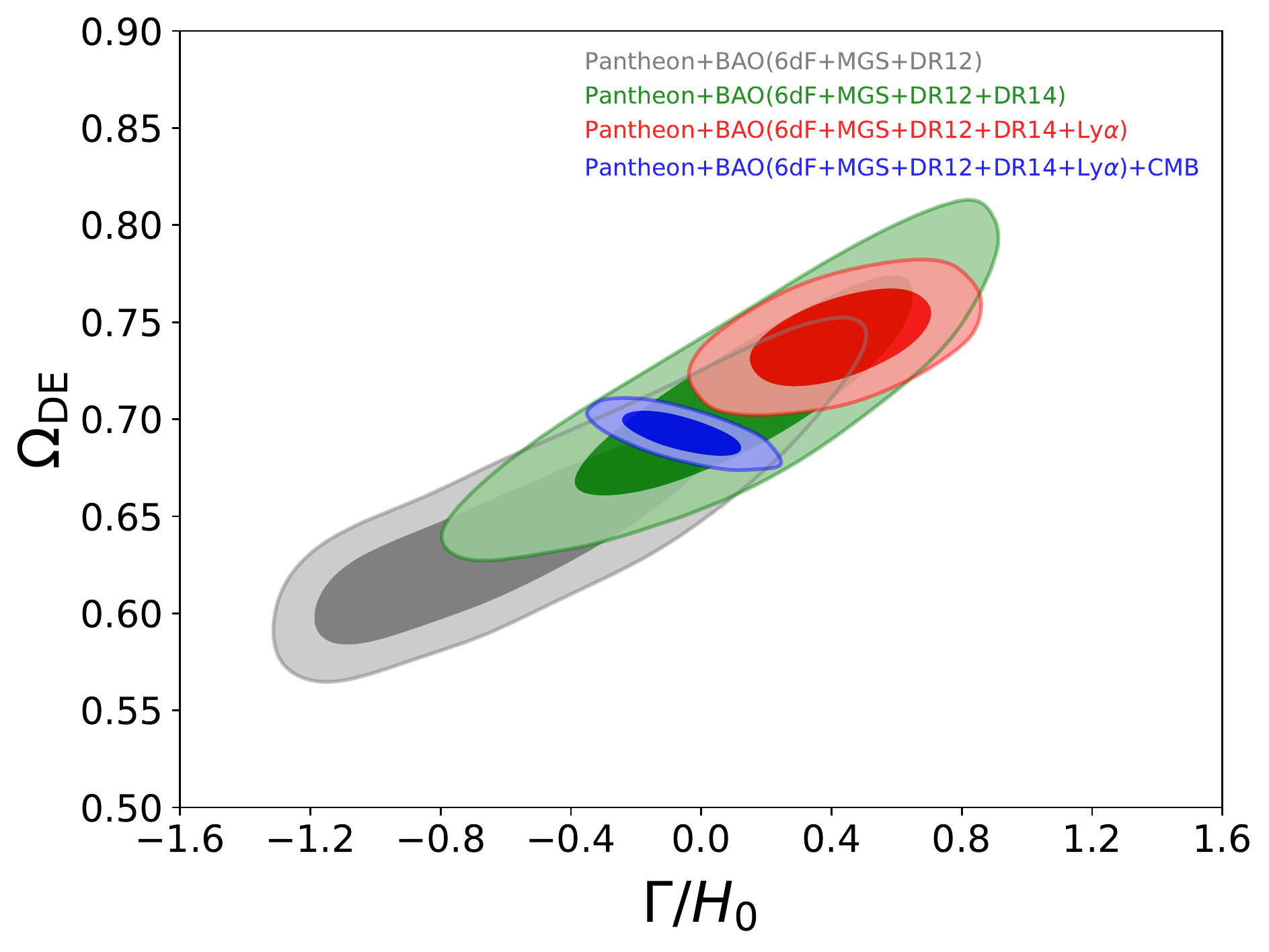}
\caption{Observed constrains on model I. The upper two plots show the marginalized 1D likelihood for $\Omega_{m,0}h^2$ (left) and $\Gamma/H_0$ (right). The lower two plots show the marginalized 1$\sigma$ and 2$\sigma$ contours for matter density vs Hubble constant (left) and $\Omega_{\rm{DE}}$ vs $\Gamma/H_0$ (right). Different color denotes for the constraint results from different data sets. }\label{fig:res_m1}
\end{figure*}

\begin{figure*}
\centering
\includegraphics[width=1\textwidth]{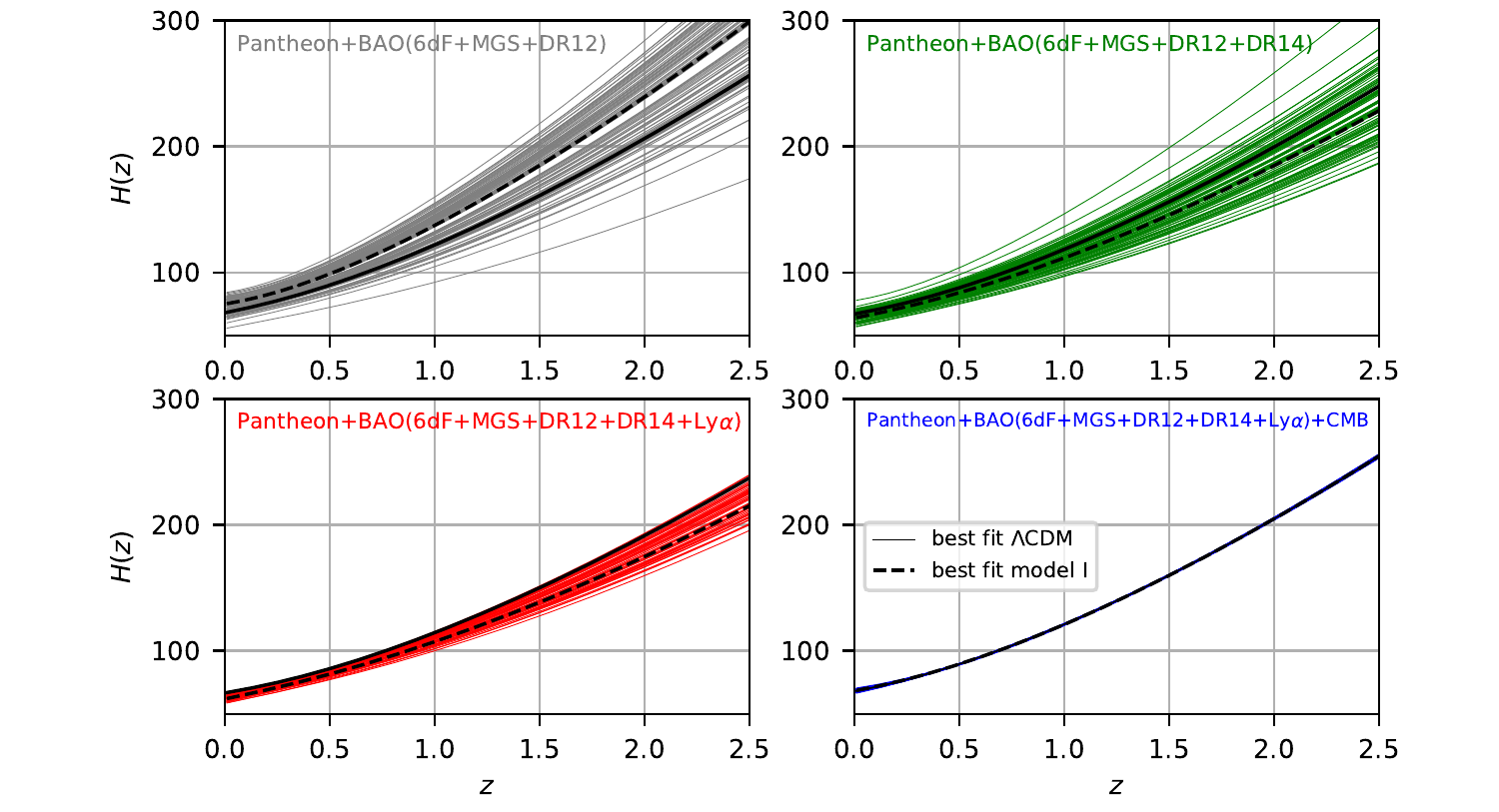}
\caption{The Hubble parameter $H(z)$ for the metastable DE model I obtained with different data combinations. The solid black lines and the dashed black lines show $H(z)$ from the best fit of the  $\rm{\Lambda}$CDM model and the metastable DE model I with same data set, respectively.}\label{fig:Hz_m1}
\end{figure*}

\begin{figure*}
\centering
\includegraphics[width=1\textwidth]{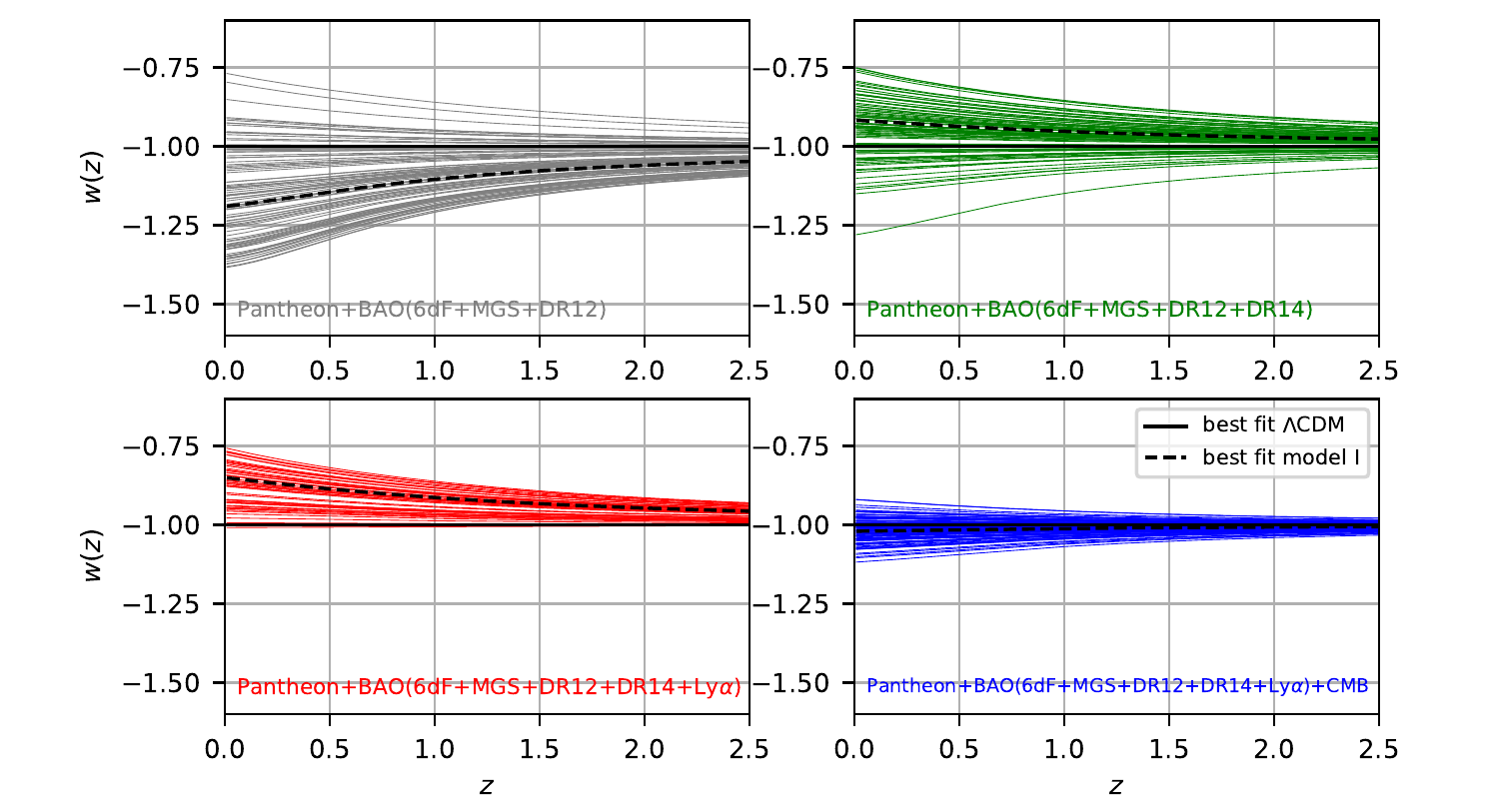}
\caption{The equation of state of dark energy as a function of redshift for the metastable DE model I obtained with different data combinations.  The solid black lines and the dashed black lines show $w(z)$ from the best fit of the $\rm{\Lambda}$CDM model and the metastable DE model I with same data set, respectively.}\label{fig:wz_m1}
\end{figure*}

\begin{figure*}
\centering
\includegraphics[width=1\textwidth]{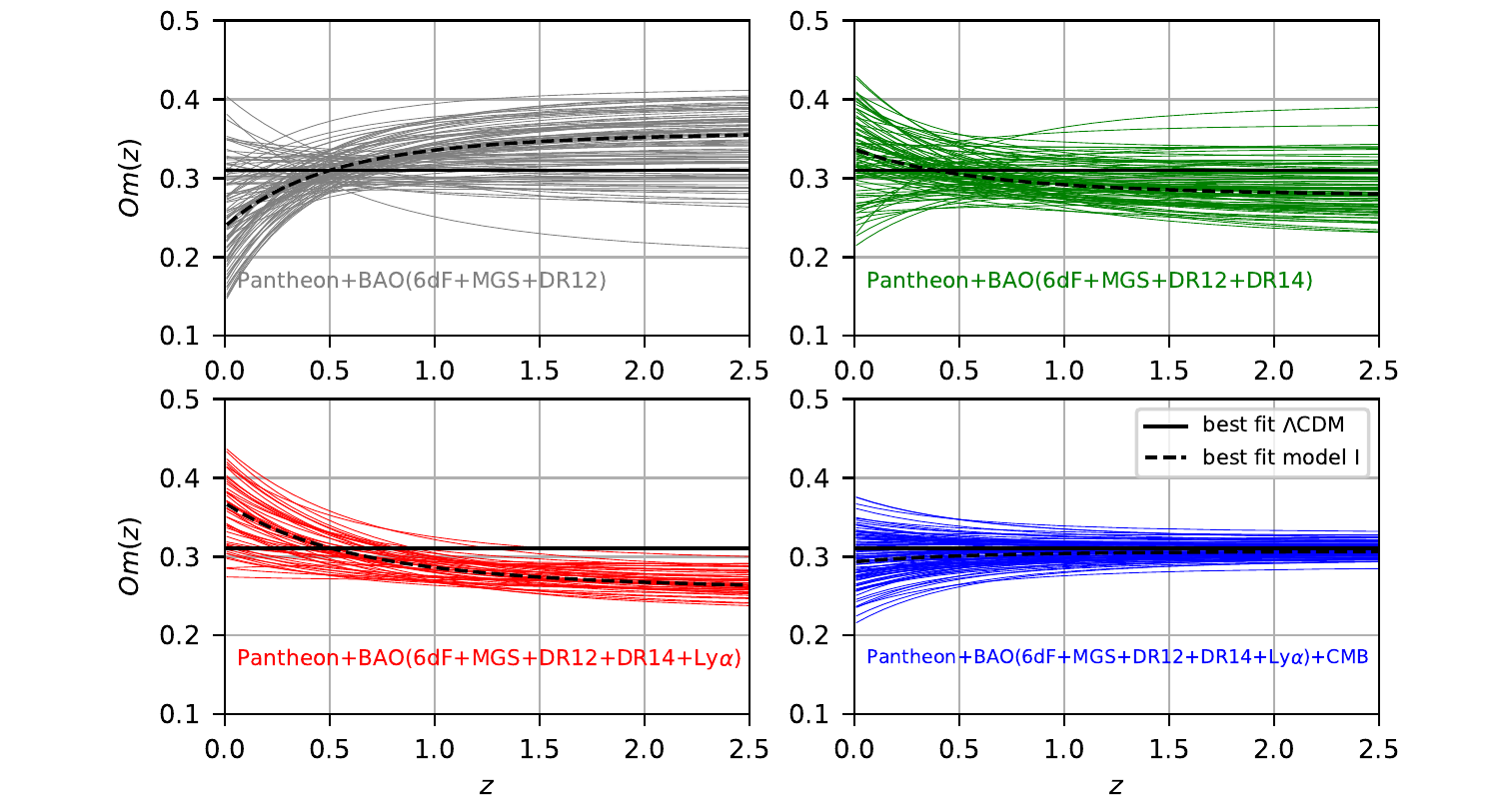}
\caption{The $Om$ diagnostic as a function of redshift for the metastable DE model I obtained with different data combinations.  The solid black lines and the dashed black lines show $Om(z)$ from the best fit of the $\rm{\Lambda}$CDM model and the metastable DE model I with the same data set, respectively.}\label{fig:Omz_m1}
\end{figure*}

\begin{figure*}
\centering
\includegraphics[width=1\textwidth]{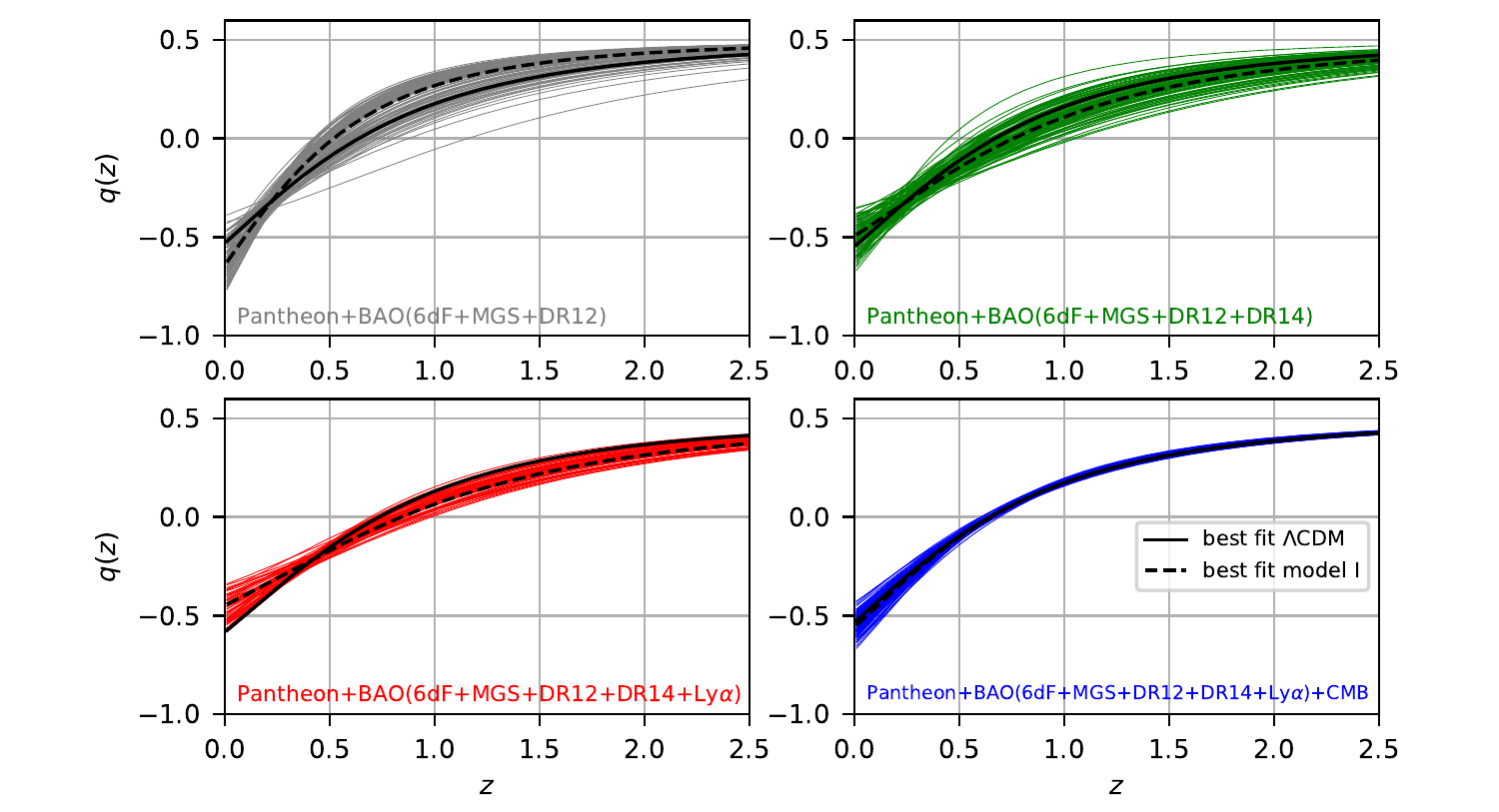}
\caption{The deceleration parameter as a function of redshift for the metastable DE model I obtained with different data combinations.  The solid black lines and the dashed black lines show $q(z)$ from the best fit of the $\rm{\Lambda}$CDM model and the metastable DE model I with the same data set, respectively.}\label{fig:qz_m1}
\end{figure*}

\begin{table*}[!t]
\centering
\caption{ The best fit of cosmological parameters (the first row in each parameter row) for the metastable DE model I and its mean value together with its marginalized 1$\sigma$ uncertainties  (the second row in each parameter row) as well as their $\chi^2$ value. } \label{tab:res_m1}
\begin{tabular}{ccccc}
\hline
\hline 
data       & Pantheon           & Pantheon            & Pantheon             & Pantheon \\
           & +BAO(6dF+MGS+DR12) & +BAO(6dF+MGS+DR12   & +BAO(6dF+MGS+DR12    &+BAO(6dF+MGS+DR12 \\
parameters &                    & +DR14)              &  +DR14+Ly$\alpha$)   & +DR14+Ly$\alpha$)+CMB \\
 \hline
 $\Omega_{m,0}$ & $0.360$                    & $0.276$                  & $0.256$                   &  $0.307$        \\
                & $0.360^{+0.033}_{-0.042}$  & $0.288^{+0.037}_{-0.38}$ & $0.259^{+0.017}_{-0.016}$ &  $0.307^{+0.008}_{-0.008}$        \\
 \hline
 $H_0$          & $75.09$                  & $64.03$                 & $61.98$                 &                  $68.45$                  \\
                & $75.01^{+4.71}_{-5.80}$  & $65.53^{+4.60}_{-4.41}$ & $62.29^{+1.99}_{-1.94}$ &                  $68.38^{+0.83}_{-0.82}$        \\
  \hline
 $\Omega_{m,0}h^2$ & $0.203 $                   & $0.113$                   & $0.098$                    &                  $0.144$        \\
                   & $0.203^{+0.046}_{-0.050}$  & $0.124^{+0.036}_{-0.030}$ & $0.100^{+0.0124}_{-0.190}$ &                  $0.144^{+0.001}_{-0.001}$        \\
 \hline
 $\Gamma/H_0$   & $-0.57$                  & $0.25$                 & $0.45$                 &                  $-0.06$        \\
                & $-0.55^{+0.46}_{-0.40}$  & $0.14^{+0.34}_{-0.37}$ & $0.42^{+0.19}_{-0.19}$ &                  $-0.06^{+0.13}_{-0.12}$        \\
 \hline
 $\chi^2$       & 1041.52            & 1046.42 &  1058.94  & 1070.94 \\
\hline
\hline
\end{tabular}
\end{table*}

In Fig.~\ref{fig:res_m1} we show the results for the metastable DE model I, in which DE decays exponentially. We show the 1D likelihoods for $\Omega_{m,0}h^2$ and $\Gamma/H_0$ in the upper plots and the 2D marginalized 1$\sigma$ and 2$\sigma$ contours in the lower plots. As before, different colors imply different data combinations. The two left plots should be compared with Fig.~\ref{fig:res_lcdm}. The best fit for the cosmological parameters of model I and the marginalized 1$\sigma$ uncertainties as well as the $\chi^2$ of each data combination are presented in Table~\ref{tab:res_m1}. Compared with $\rm{\Lambda}$CDM, the confidence contours are much larger. However, the $H_0$ tension between higher redshift BAO measurements from Ly$\alpha$ and CMB increases. Lower matter density and lower Hubble parameter are favoured by adding high redshift BAO measurements from Ly$\alpha$. 

Moreover, from the two right plots of Fig.~\ref{fig:res_m1} we can see that the constraint on $\Gamma/H_0$ obtained with Pantheon+BAO(6dF+MGS+DR12) (grey curves) support $\Gamma<0$ while the results obtained with Pantheon+BAO(6dF+MGS+DR12+DR14) (green curves) support either $\Gamma\,<\,0$ or $\Gamma\,>\,0$, which means that Pantheon+BAO(6dF+MGS+DR12) data suggest that the DE density is increasing, while for Pantheon in combination with BAO(6dF+MGS+DR12+DR14) the best fits on $\Gamma$  don't show any preference for the DE density to be either increasing or decreasing. However, adding Ly$\alpha$ BAO data into the analysis gives $\Gamma \,>\,0$ (red curves), which means DE density decays with time, in other words the DE density at earlier times was larger than at present.  On the other hand,  results obtained after including the CMB distance prior lie close to  $\Gamma\,=\,0$ (blue curves) suggesting that $\rm{\Lambda}$CDM is preferred by the CMB data set. This might be due to the fact that CMB distance priors are obtained assuming $\rm{\Lambda}$CDM cosmology. 

In Fig.~\ref{fig:Hz_m1}, Fig.~\ref{fig:wz_m1}, Fig.~\ref{fig:Omz_m1} and Fig.~\ref{fig:qz_m1}, we show the Hubble parameter as a function of redshift $H(z)$, the equation of state of dark energy as a function of redshift $w(z)$, the $Om$ diagnostic $Om(z)\,=\,(h^2(z)-1)/[(1+z)^3-1]$ and the deceleration parameter $q(z)\,=\,-\dot{H}/H^2-1$ for the metastable DE model I. We plot 100 samples for these parameters randomly chosen from within 2$\sigma$ range of the MCMC chains corresponding to different data sets.

\begin{figure*}
\centering
\includegraphics[width=0.45\textwidth]{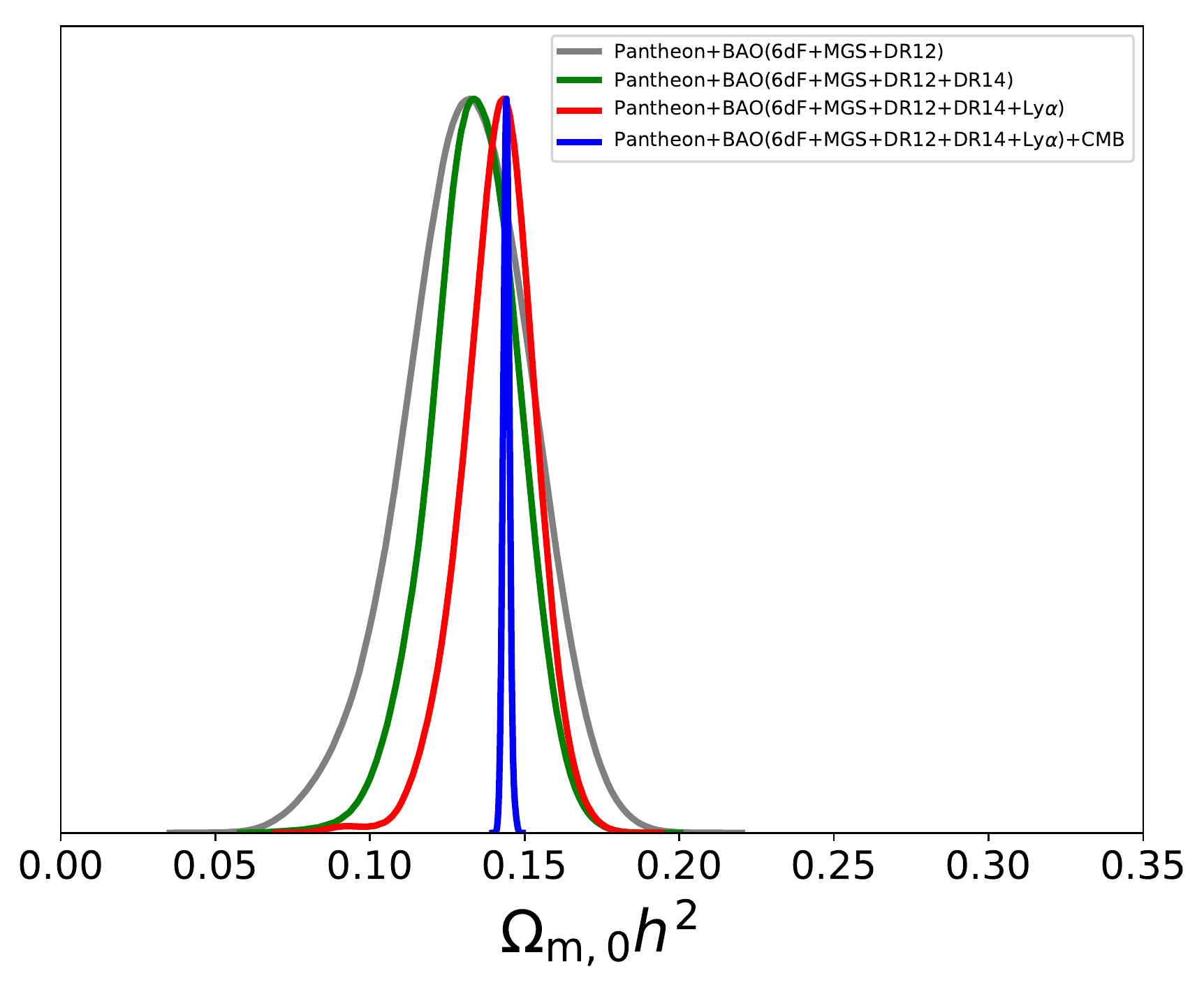}
\includegraphics[width=0.45\textwidth]{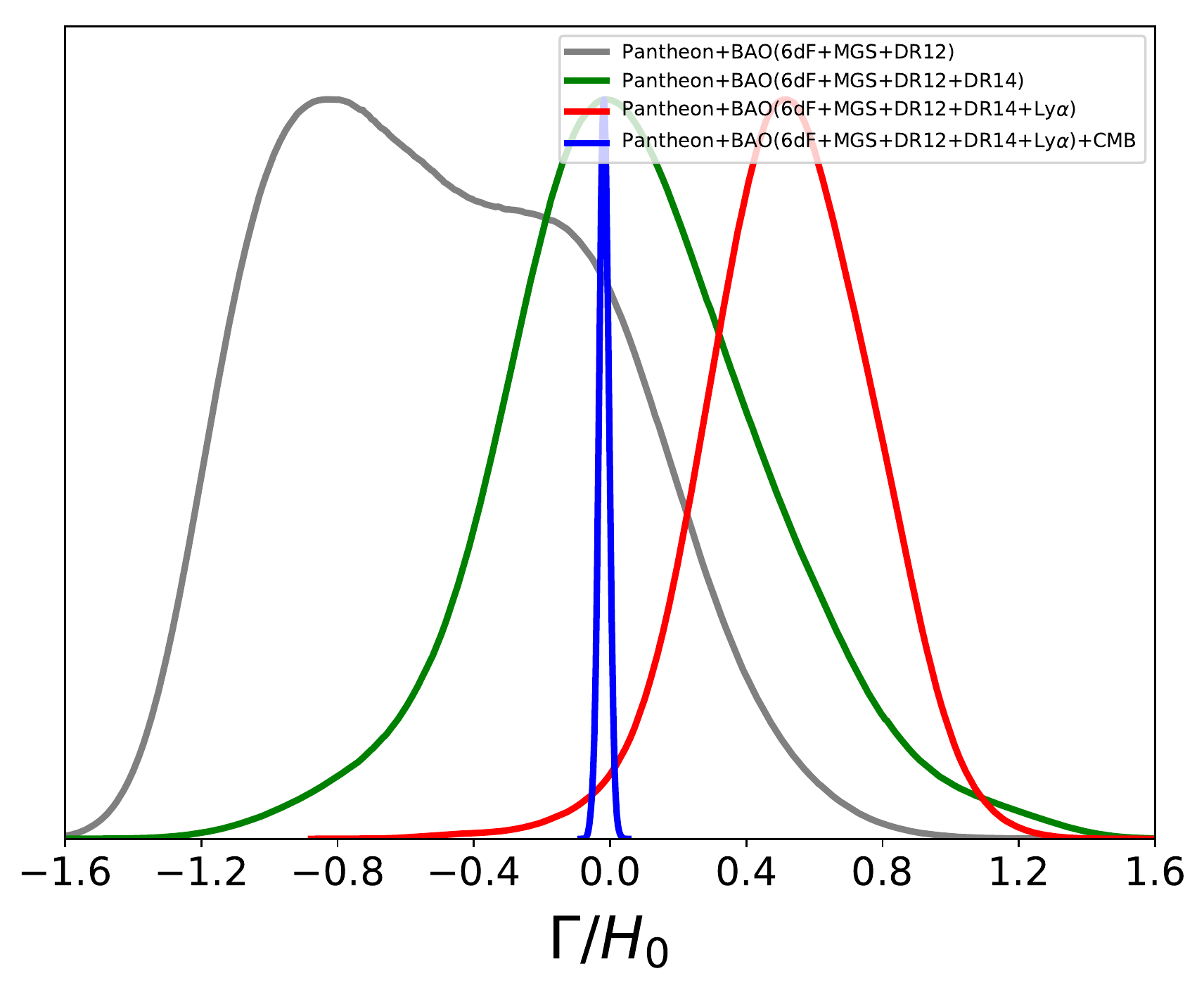}
\includegraphics[width=0.47\textwidth]{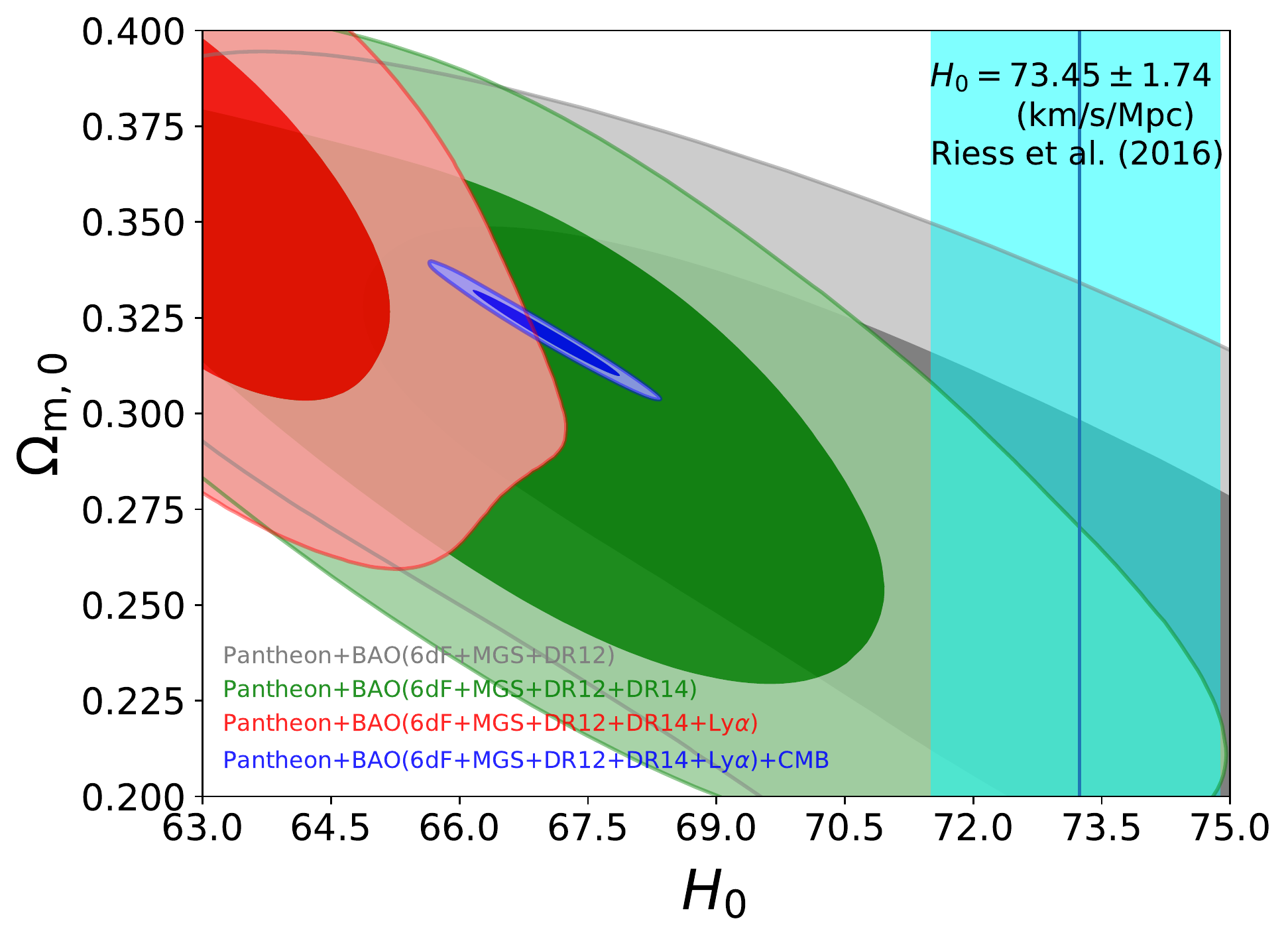}
\includegraphics[width=0.45\textwidth]{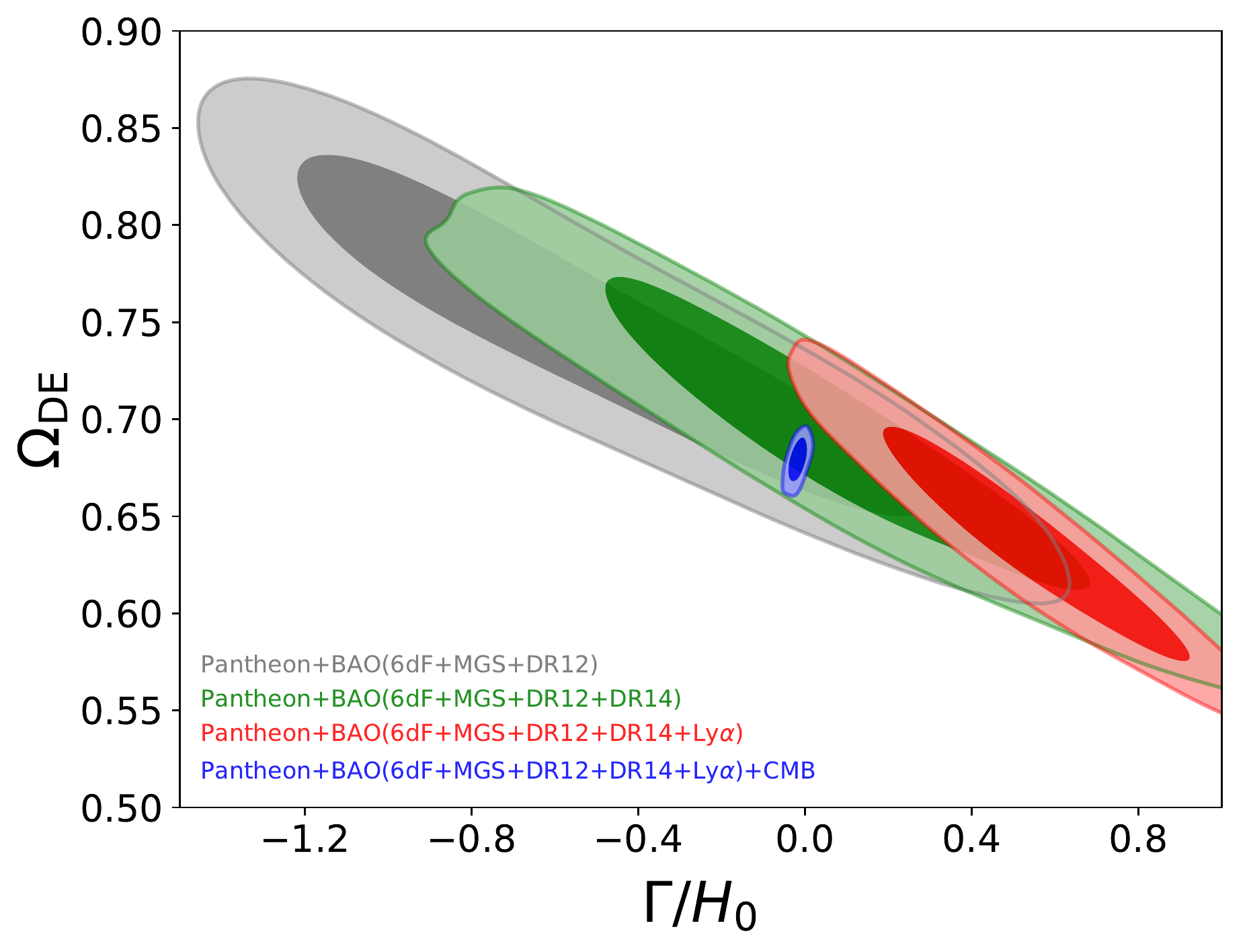}
\caption{The constrain results for Model II. The upper two plots show the marginalized 1D likelihood for $\Omega_{m,0}h^2$ (left) and $\Gamma/H_0$ (right). The lower two plots show the marginalized 1$\sigma$ and 2$\sigma$ regions for matter density vs Hubble constant (left) and $\Omega_{\rm{DE}}$ vs $\Gamma/H_0$ (right). Different color denotes for the constraint results from different data sets.}\label{fig:res_m2}
\end{figure*}

\begin{table*}[!t]
\centering
\caption{ The best fit of cosmological parameters (the first row in each parameter row) and its mean value together with its marginalized 1$\sigma$ uncertainties (the second row in each parameter row) for metastable DE model II obtained from different data combination. The last row show the $\chi^2$ value of each data combination. } \label{tab:res_m2}
\begin{tabular}{ccccc}
\hline
\hline 
data       & Pantheon           & Pantheon            & Pantheon             & Pantheon \\
           & +BAO(6dF+MGS+DR12) & +BAO(6dF+MGS+DR12   & +BAO(6dF+MGS+DR12    &+BAO(6dF+MGS+DR12 \\
parameters &                    & +DR14)              &  +DR14+Ly$\alpha$)   & +DR14+Ly$\alpha$)+CMB \\
 \hline
 $\Omega_{m,0}$ & $0.253$                    & $0.314$                    & $0.367$                   &                  $0.319         $        \\
                & $0.263^{+0.056}_{-0.060}$  & $0.304^{+0.048}_{-0.048}$ & $0.362^{+0.042}_{-0.043}$ &                  $0.321^{+0.008}_{-0.007}$        \\
 \hline
 $H_0$ & $72.61$                    & $66.74$                  &   $62.30$                  &              $67.11$        \\
       & $71.80^{+4.71}_{-4.58}$    & $66.94^{+2.55}_{-3.43}$ &    $62.64^{+1.85}_{-1.70}$  &              $66.98^{+0.56}_{-0.57}$        \\
  \hline
 $\Omega_{m,0}h^2$ & $0.133$                    & $0.137$                   & $0.142$                   &                  $0.144$        \\
                   & $0.134^{+0.019}_{-0.021}$  & $0.136^{+0.014}_{-0.014}$ & $0.142^{+0.012}_{-0.013}$ &                  $0.144^{+0.001}_{-0.001}$        \\
\hline
 $\Gamma/H_0$   & $-0.78$                  & $0.03$                 & $0.57$                   &                  $-0.01$        \\
                & $-0.47^{+0.50}_{-0.50}$  & $0.07^{+0.40}_{-0.30}$ & $0.51^{+0.27}_{-0.25}$ &                  $-0.02^{+0.01}_{-0.01}$        \\
 \hline
 $\chi^2$ & 1041.55  & 1046.50 &  1059.08  & 1071.03 \\
\hline
\hline
\end{tabular}
\end{table*}

\begin{figure*}
\centering
\includegraphics[width=1\textwidth]{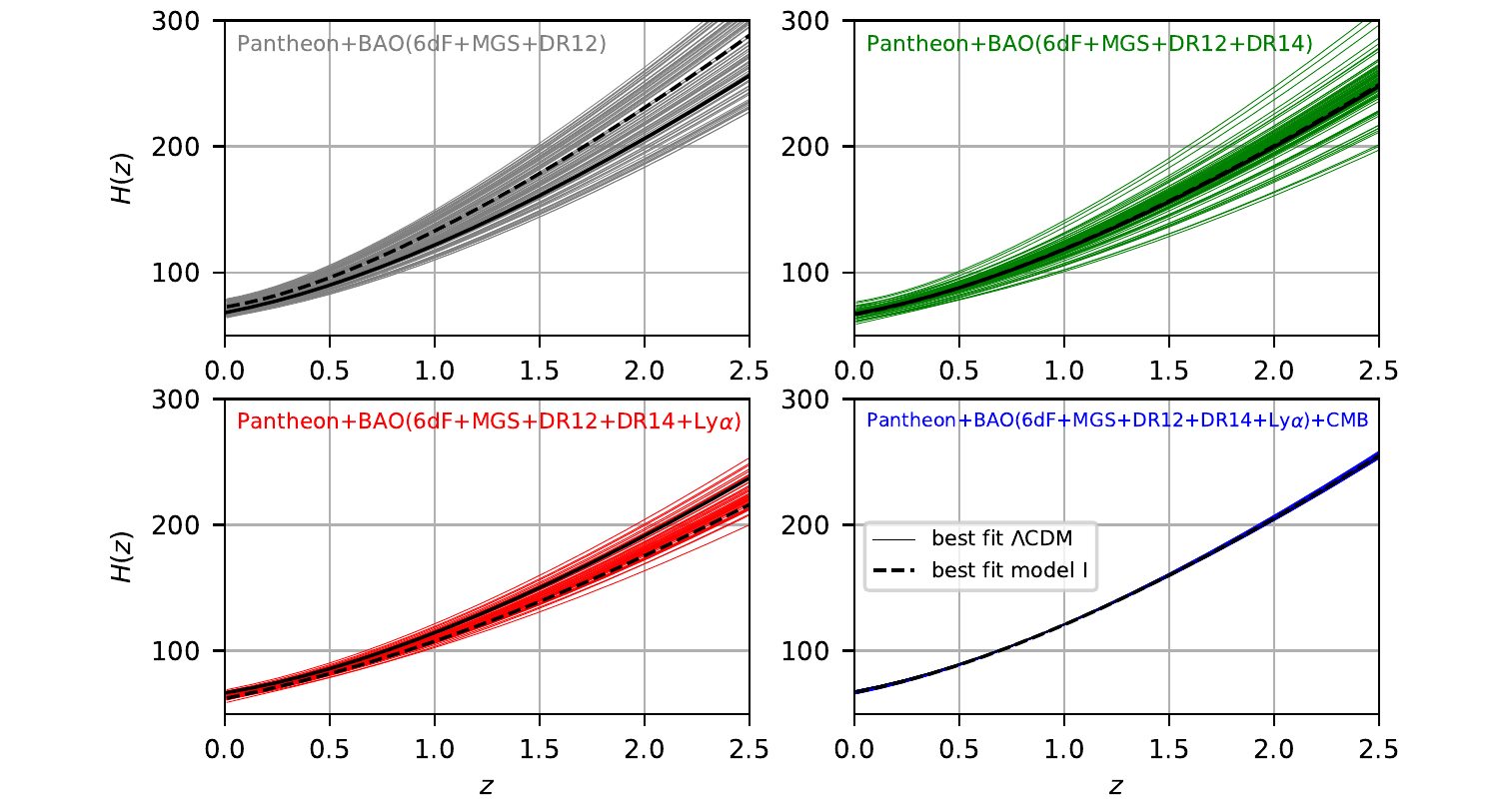}
\caption{The Hubble parameter $H(z)$ for the metastable DE model II obtained with different data combinations. The solid black lines and the dashed black lines show $H(z)$ from the best fit of the $\rm{\Lambda}$CDM model and the metastable DE model II with the same data set, respectively.}\label{fig:Hz_m2}
\end{figure*}

\begin{figure*}
\centering
\includegraphics[width=1\textwidth]{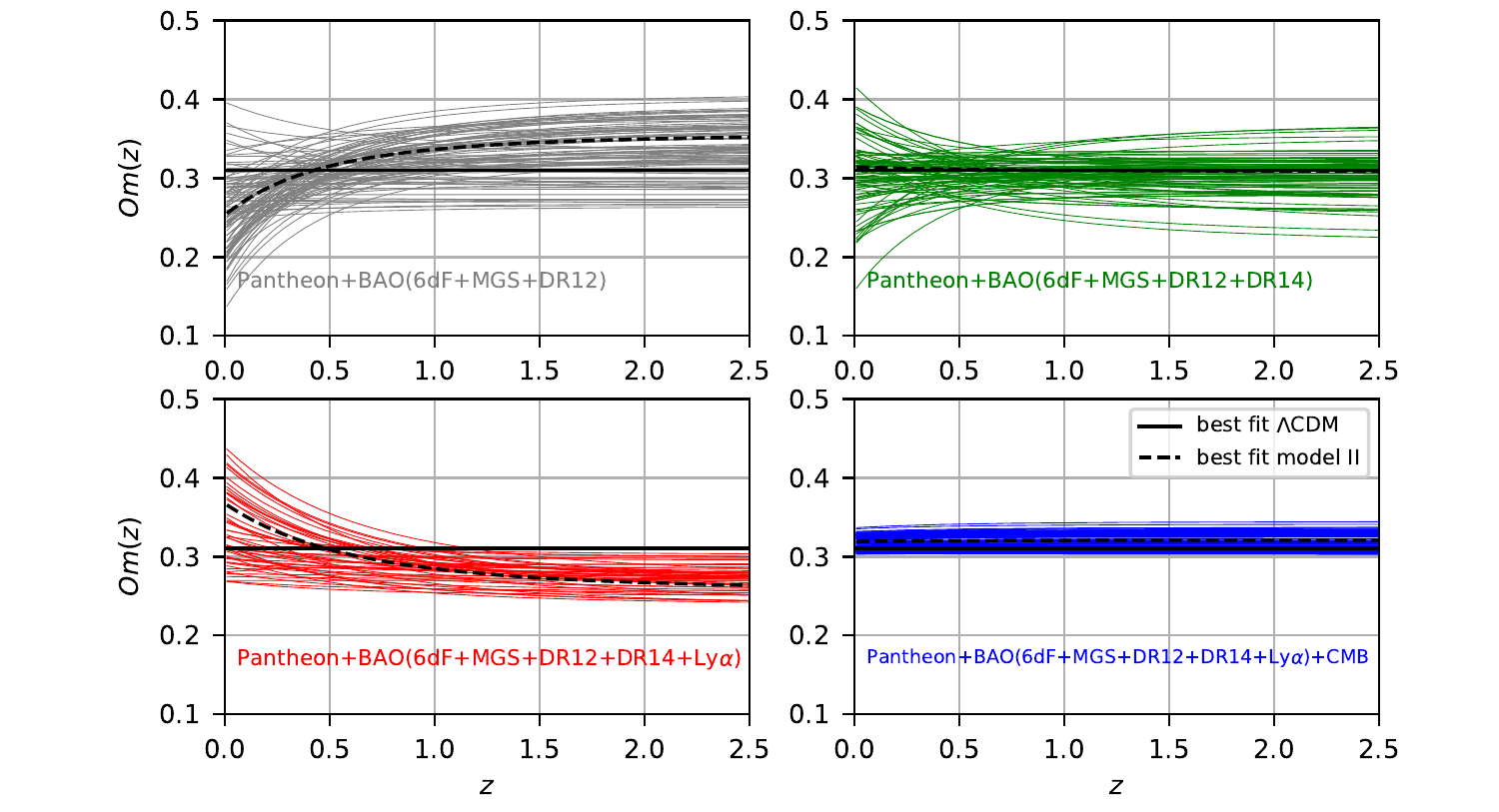}
\caption{The $Om$ diagnostic as a function of redshift for the metastable DE model II obtained with different data combinations. The solid black lines and the dashed black lines show $Om(z)$ from the best fit of the $\rm{\Lambda}$CDM model and the metastable DE model II with the same data set, respectively.}\label{fig:Omz_m2}
\end{figure*}

\begin{figure*}
\centering
\includegraphics[width=1\textwidth]{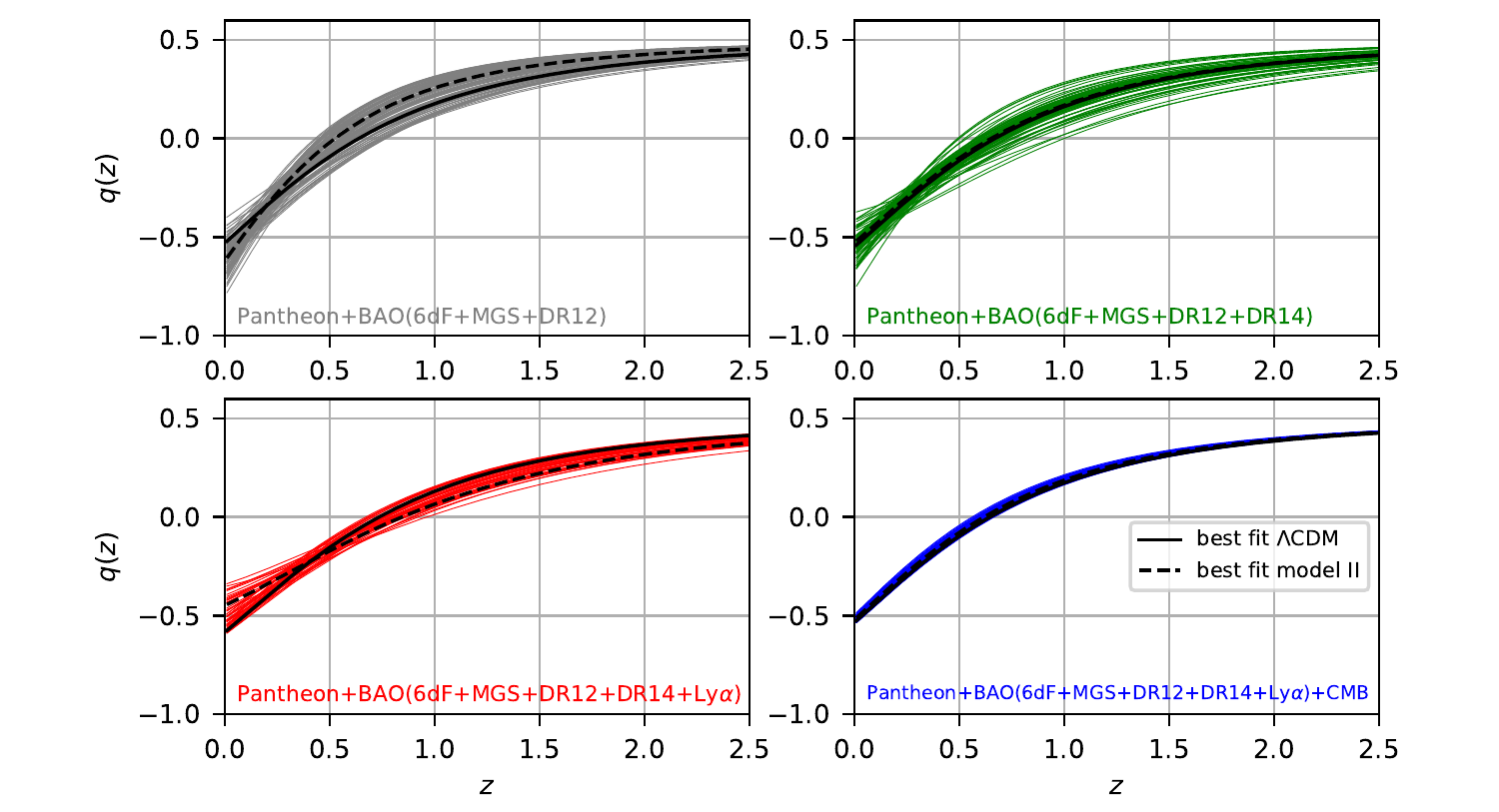}
\caption{The deceleration parameter as a function of redshift for the metastable DE model II obtained with different data combinations. The solid black lines and the dashed black lines show $q(z)$ from the best fit of the $\rm{\Lambda}$CDM model and the metastable DE model II with the same data set, respectively.}\label{fig:qz_m2}
\end{figure*}

\vspace{0.5cm}
 Fig.~\ref{fig:res_m2} shows the corresponding results for the metastable DE model II. The upper two plots show the 1D likelihoods for $\Omega_{\rm{m,0}}h^2$ (left) and $\Gamma/H_0$ (right) obtained from different data combinations. The lower plots show the 2D marginalized 1$\sigma$ and 2$\sigma$ regions for $\Omega_{\rm{m,0}}$ vs $H_0$ (left plot) and $\Omega_{\rm{DE}}$ vs $\Gamma/H_0$ (right plot). The details of the best fits and 1$\sigma$ uncertainties for parameters of model II are summarized in Table~\ref{tab:res_m2}. From the left bottom plot and Table~\ref{tab:res_m2}, we can see that adding BAO measurement from Ly$\alpha$ makes the best fit of $\Omega_{\rm{m,0}}$ larger than it obtained without BAO measurement from Ly$\alpha$. While the constrain results for $H_0$ become lower when including BAO measurement from Ly$\alpha$. However, adding CMB distance prior to the data set pushes the results back to higher $H_0$ and lower $\Omega_{\rm{m,0}}$. The $H_0$ tension still exists between CMB and BAO measurement from Ly$\alpha$. However, as can be seen from the upper left plot, $\Omega_{\rm{m,0}}h^2$ agrees well between CMB and BAO measurement from Ly$\alpha$ since the degeneracy of contours for $\Omega_{\rm{m,0}}$ and $H_0$ changes.

Now let's look at the two right plots, which focus on the constraints on $\Gamma$ from observations. As mentioned earlier, $\Gamma\,>\,0$ implies that DE decays into dark matter, while $\Gamma\,<\,0$ means the opposite: dark matter decays into DE. Our model becomes $\rm{\Lambda}$CDM when $\Gamma\,=\,0$. We find that, with Pantheon in combination with BAO data from 6dFGS, MGS and BOSS DR12, the best fit for $\Gamma$ supports the transfer of energy from dark matter to dark energy, while adding BAO measurements from eBOSS DR14 and Ly$\alpha$ favours the opposite. Including CMB distance prior to the analysis gives $\Gamma\,\simeq\,0$, which means that the DE energy density remains unchanged. 

In Fig.~\ref{fig:Hz_m2}, Fig.~\ref{fig:Omz_m2} and Fig.~\ref{fig:qz_m2}, we also show the Hubble parameter as a function of redshift $H(z)$, the $Om$ diagnostic $Om(z)\,=\,(h^2(z)-1)/[(1+z)^3-1]$ and the deceleration parameter $q(z)\,=\,-\dot{H}/H^2-1$ for the metastable DE model II, respectively. We plot 100 samples for these parameters randomly chosen from within 2$\sigma$ range of the MCMC chains corresponding to different data sets.

\vspace{0.5cm}
\begin{figure*}
\centering
\includegraphics[width=0.32\textwidth]{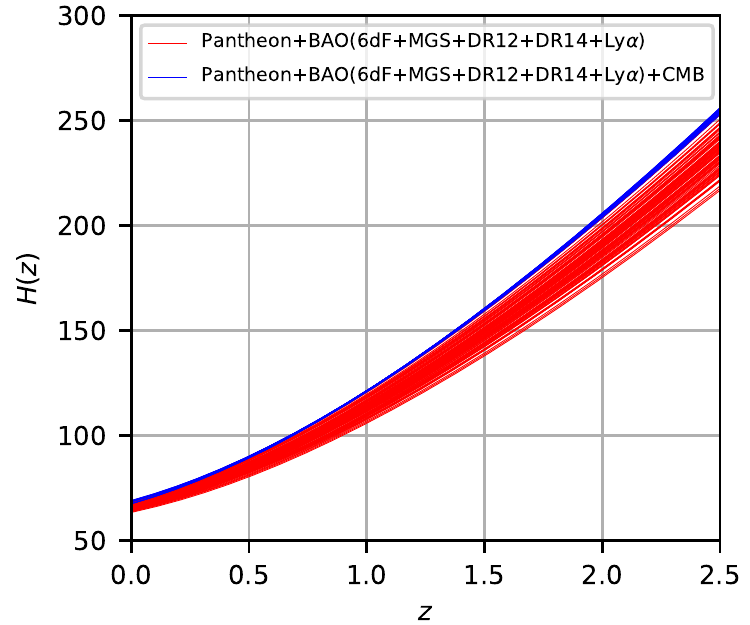}
\includegraphics[width=0.32\textwidth]{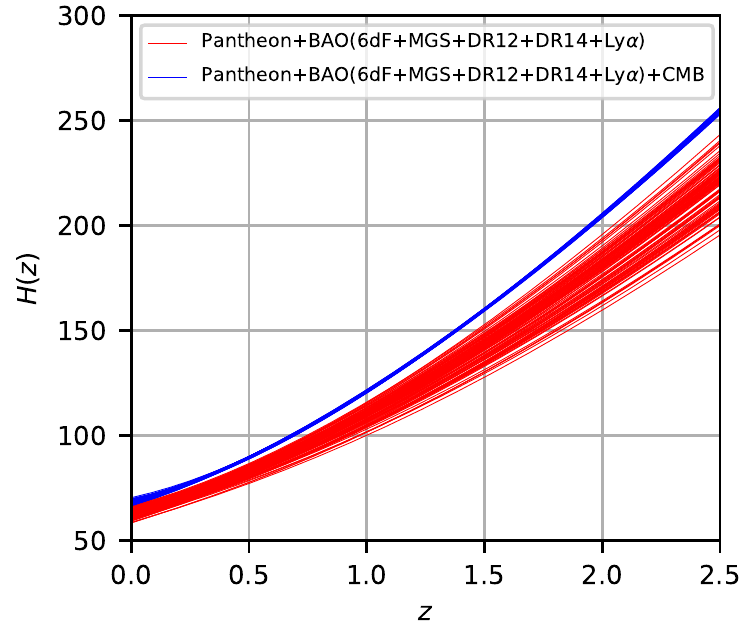}
\includegraphics[width=0.32\textwidth]{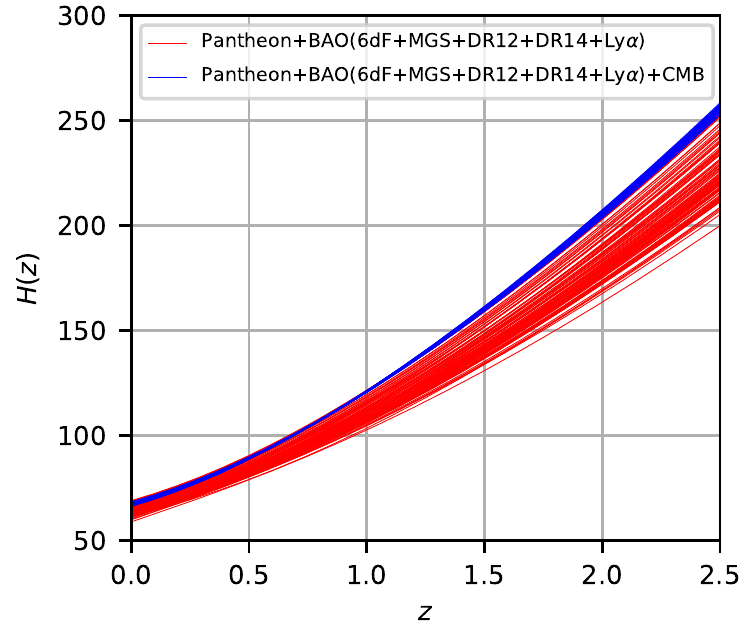}
\caption{Hubble parameter for different cosmological models constrained with two different data sets described above.{{ From left to right, we show Hubble parameter as a function of redshift for standard $\Lambda$CDM model, metastable model I and model II, respectively.}}}\label{fig:Hz_com}
\end{figure*}

In Fig.~\ref{fig:Hz_com}, we show the $H(z)$ samples within 2$\sigma$ confidence level for different cosmological models obtained with different data combinations including SNe Ia, BAO and CMB. The left plot shows the $\rm{\Lambda}$CDM model results which shows that for the data set combination with and without CMB data the sample shows little overlap. While for the metastable DE model I, which is shown in the middle plot of Fig.~\ref{fig:Hz_com}, there is apparently no overlap. The results for the metastable DE model II are shown in the right plot of Fig.~\ref{fig:Hz_com}, which is similar to $\rm{\Lambda}$CDM.

Using current data sets of SNe Ia, BAO and CMB, we find that the $H_0$ tension between CMB and BAO measurements from Ly$\alpha$ existing in $\rm{\Lambda}$CDM model (see Fig.~\ref{fig:res_lcdm}) become larger when compared with the results obtained from  previous data sets (see Fig.~1 in \citet{shafieloo2017metastable}) and metastable DE models cannot reduce this $H_0$ tension. Including the CMB distance prior to our analysis shows that both model I and model II are consistent with $\rm{\Lambda}$CDM model, while without CMB data, the results from Pantheon in combination with BAO(6dF+MGS+DR12+DR14+Ly$\alpha$) support that DE density is decaying (exponentially for the model I and into dark matter density for the model II). 

\section{summary}\label{sec:sum}
In this work, we revisit two metastable DE models proposed in \citet{shafieloo2017metastable} confronting them with the Pantheon SNe Ia sample, BAO measurements derived from 6dFGS, the SDSS DR7 MGS sample, the BOSS DR12, the eBOSS DR14 and high redshift BAO measurement from the Ly$\alpha$ forest in combination with the CMB distance prior from the final \textit{Planck} release in 2018. 

%The metastable DE models was put forward in \citet{shafieloo2017metastable} in order to alleviate the $H_0$ tension between CMB measurements and high redshift BAO measurements from Ly$\alpha$ existing in $\rm{\Lambda}$CDM model. 

In the metastable DE models, the DE density decays exponentially in the model I and decays into dark matter in the model II (the reverse process DM $\to$ DE is also
permitted). The specific feature of these two models is that the decay rate is a constant and depends only on intrinsic properties of dark energy
and not on other factors such as cosmological expansion, etc.

We estimate some key cosmological parameters assuming standard $\Lambda$CDM on the one hand, and the two metastable DE models on the other. We find that with current data sets, the $\Omega_{0m}h^2$ tension  between CMB and high redshift BAO measurement from Ly$\alpha$ becomes significant in $\Lambda$CDM  and also in the model I. The model II shows slightly better consistency. We should note that the degeneracy direction for $\Omega_{\rm{m,0}}$ vs $H_0$ for the model II is different from the model I, that makes constraints on the derived parameter $\Omega_{\rm{m,0}}h^2$ to agree better with CMB and high redshift BAO measurements. Marginalised probability distribution function for $H_0$ in the metastable dark energy models including supernovae and all BAO data (except the Ly$\alpha$ BAO data) shows clear consistency with the results including Planck CMB constraints. However, including Ly$\alpha$ BAO data (and without Planck CMB measurements) changes the constraints on $H_0$ dramatically, lowering it to a central value of $62\,$km/sec/Mpc. Since local measurements of the Hubble constant place its value to be around $73\,$km/sec/Mpc, the situation seems to be very conflicting. In fact Ly$\alpha$ BAO data, Planck CMB data and the local measurement of $H_0$, each pull our models to a different region in parameter space. This could be due to tension between different data sets. Possible resolutions of this dilemma might lie in systematics in some of the data, a more complicated form of the expansion history (which needs to be reconstructed carefully to satisfy all observations) or an unconventional model of the early Universe \citep{hazra2019parameter}.   

{{We should note here that from a statistical point of view and to compare the analysed metastable dark energy models in this paper with $\rm{\Lambda}$CDM model, we do not expect that these models perform better than the standard $\rm{\Lambda}$CDM model by estimating the Bayes factor (as done in \citet{pan2019reconciling}, analysing the model proposed in \citet{li2019phenomenologically}). The fact is that having an extra degree of freedom in these metastable models, we have not achieved a substantial improvements in the likelihood estimations (as shown in Table~\ref{tab:res_lcdm},~\ref{tab:res_m1},~\ref{tab:res_m2}). In such a situation, the Bayesian analysis prefers a model with lower degrees of freedom ($\rm{\Lambda}$CDM model). However, future data with higher precision and better control of systematics might change the current situation and we might get substantially different likelihoods for these models in comparison to the standard $\rm{\Lambda}$CDM model. This might not be surprising as the standard $\rm{\Lambda}$CDM model has problems fitting the low and high redshift data simultaneously, and with higher precision data the likelihood for this model might get substantially worsen. This is the main reason why we should continue to study models that might not be currently favoured compared to the standard model while they may have interesting phenomenological or theoretical properties.}}

By the time we were finalising our work, a recent work by \citet{riess2019large} came out and in their work they showed a larger $H_0$ tension between locally measurement and the value inferred from \textit{Planck} CMB and $\rm{\Lambda}$CDM. It is therefore extremely important to understand the nature of these tensions.

%$\rm{\Lambda}$CDM model. 

%Similar to the results in \citet{shafieloo2017metastable}, metastable DE model II is able to reduce the $H_0$ tension. Secondly, adding BAO measurement from Ly$\alpha$ gives $\Gamma\,>\,0$, which means DE density is decreasing. Including CMB measurements into the analysis gives $\Gamma\,\sim\,0$, which shows consistence with $\rm{\Lambda}$CDM model. Thirdly, the degeneracy for $\Omega_{\rm{m,0}}$ vs $H_0$ for model II is different from model I, which makes constraints on the derived parameter, $\Omega_{\rm{m,0}}h^2$ agrees well between CMB and high redshift BAO measurements.

\acknowledgments
X. Li thanks Ryan E. Keeley and Benjamin L'Huillier for valuable suggestions. X. Li thanks Jingzhao Qi for kindly instructions about \texttt{Cosmomc}. X. Li and A.S. would like to acknowledge the support of the National Research Foundation of Korea (NRF-2016R1C1B2016478). X. Li was supported by the fund of Hebei Normal University. X. Li was supported by the Strategic Priority Research Program of the Chinese Academy of Sciences, Grant No. XDB23000000. A.A.S. was partly supported by the program KP19-270 "Questions of the origin and evolution of the Universe" of the Presidium of the Russian Academy of Sciences. This work benefits from the high performance computing clusters Polaris and Seondeok at the Korea Astronomy and Space Science Institute.

\bibliography{ref}

\end{document}